\documentclass[longauth]{aa}

\usepackage{graphicx}
\usepackage{txfonts}
\usepackage{lipsum}
\usepackage{lscape}
\usepackage{placeins}
\usepackage{natbib}  
\usepackage{hyperref}
\hypersetup{colorlinks=true, linkcolor=blue, citecolor=blue, filecolor=blue, urlcolor=blue}
\bibpunct{(}{)}{;}{a}{}{,}
\usepackage{afterpage}
\usepackage{siunitx}
\usepackage{array}
\usepackage[normalem]{ulem}
\usepackage[varvw]{newtxmath}
\usepackage{stfloats}
\DeclareUnicodeCharacter{2212}{-}
\usepackage{xspace}
\usepackage[dvipsnames]{xcolor}
\usepackage{cancel}
\usepackage{bm}
\usepackage{float}
\FloatBarrier
\usepackage{shortcuts}

\begin{document}
    
    \title{COSMOS-Web: does halo mass alone shape the clustering of star-forming and quiescent galaxies?}
    
    \author{
        L.~Paquereau\inst{\ref{Chalmers}}
        \and C.~Laigle\inst{\ref{IAP}}
        \and H.~J.~McCracken\inst{\ref{IAP}}
        \and O.~Ilbert \inst{\ref{LAM}}
        \and H.~B.~Akins\inst{\ref{UAT}}
        \and R.~Arango-Togo \inst{\ref{IAP}}
        \and N.~Bình \inst{\ref{seattle}}
        \and C.~M.~Casey\inst{\ref{SantaB},\ref{DAWN}}
        \and Y.~Dubois\inst{\ref{IAP}}
        \and M.~Franco\inst{\ref{CEA}}
        \and G.~Gozaliasl \inst{\ref{finland2},\ref{finland}}
        \and S.~Harish\inst{\ref{STScI},\ref{Rochester}}
        \and M.~Hirschmann\inst{\ref{EPFL},\ref{trieste}}
        \and B.~Jego\inst{\ref{Strasbourg}}
        \and A.~Kaminsky\inst{\ref{miami}}
        \and J.~S.~Kartaltepe\inst{\ref{Rochester}}
        \and A.~M.~Koekemoer\inst{\ref{STScI}}
        \and D.~Le Borgne\inst{\ref{IAP}}
        \and J.~S.~W.~Lewis\inst{\ref{CEA}}
        \and D.~Liu\inst{\ref{Nanjing}}
        \and G.~E.~Magdis \inst{\ref{DAWN},\ref{NBI},\ref{lyngby}}
        \and J.~McKinney\inst{\ref{UAT}}
        \and W.~Mercier \inst{\ref{LAM}}
        \and L.~Moscardini \inst{\ref{Bologna},\ref{INAF-Bologna},\ref{INFN_bologna}}
        \and T.~Moutard\inst{\ref{ESA}}
        \and J.~D.~Rhodes\inst{\ref{NASAJPL}}
        \and B.~E.~Robertson\inst{\ref{UnivCalifornia}}
        \and S.~Sanjaripour\inst{\ref{riverside}}
        \and M.~Shuntov\inst{\ref{DAWN},\ref{NBI}}
        \and G.~Toni\inst{\ref{Bologna},\ref{INAF-Bologna},\ref{Heidelberg}}
        \and M.~Trebitsch\inst{\ref{MeudonLUX}}
        \and L.~Tresse\inst{\ref{LAM}}
    }

   \institute{
        Department of Physics and Astronomy, Chalmers University of Technology, SE-412 96 Gothenburg, Sweden \label{Chalmers}
        \and Institut d’Astrophysique de Paris, UMR 7095, CNRS, Sorbonne Université, 98 bis boulevard Arago, F-75014 Paris, France\label{IAP}
        \and Aix Marseille University, CNRS, CNES, LAM, Marseille, France\label{LAM}
        \and Department of Astronomy, The University of Texas at Austin, 2515 Speedway Blvd Stop C1400, Austin, TX 78712, USA \label{UAT}
        \and Department of Astronomy, University of Washington, Seattle, WA 98195, USA \label{seattle}
        \and Department of Physics, University of California, Santa Barbara, Santa Barbara, CA 93106, USA\label{SantaB}
        \and Cosmic Dawn Center (DAWN), Denmark\label{DAWN}
        \and Université Paris-Saclay, Université Paris Cité, CEA, CNRS, AIM, 91191 Gif-sur-Yvette, France\label{CEA}
        \and Department of Computer Science, Aalto University, P.O. Box 15400, FI-00076 Espoo, Finland \label{finland2}
        \and Department of Physics, University of, P.O. Box 64, FI-00014 Helsinki, Finland \label{finland}
        \and Space Telescope Science Institute, 3700 San Martin Drive, Baltimore, MD 21218, USA\label{STScI}
        \and Laboratory for Multiwavelength Astrophysics, School of Physics and Astronomy, Rochester Institute of Technology, 84 Lomb Memorial Drive, Rochester, NY 14623, USA\label{Rochester}
        \and Institute of Physics, GalSpec, Ecole Polytechnique Federale de Lausanne, Observatoire de Sauverny, Chemin Pegasi 51, 1290 Versoix, Switzerland \label{EPFL}
        \and INAF, Astronomical Observatory of Trieste, Via Tiepolo 11, 34131 Trieste, Italy \label{trieste}
        \and Université de Strasbourg, CNRS, Observatoire astronomique de Strasbourg, UMR 7550, 67000 Strasbourg, France \label{Strasbourg}
        \and Department of Physics, University of Miami, Coral Gables, FL 33124, USA \label{miami}
        \and Purple Mountain Observatory, Chinese Academy of Sciences, 10 Yuanhua Road, Nanjing 210023, China \label{Nanjing}
        \and Niels Bohr Institute, University of Copenhagen, Jagtvej 128, 2200 Copenhagen, Denmark \label{NBI}
        \and DTU-Space, Technical University of Denmark, Elektrovej 327, 2800, Kgs. Lyngby, Denmark \label{lyngby}
        \and Alma Mater Studiorum University of Bologna - Department of Physics and Astronomy "Augusto Righi" (DIFA), Via Gobetti 93/2, I-40129 Bologna, Italy \label{Bologna} 
        \and INAF–Osservatorio di Astrofisica e Scienza dello Spazio, Via Gobetti 93/3, I-40129, Bologna, Italy \label{INAF-Bologna} 
        \and INFN- Sezione di Bologna, Viale Berti Pichat 6/2, I-40127 Bologna, Italy \label{INFN_bologna}
        \and European Space Agency (ESA), European Space Astronomy Centre (ESAC), Camino Bajo del Castillo s/n, 28692 Villanueva de la Cañada, Madrid, Spain \label{ESA}
        \and Jet Propulsion Laboratory, California Institute of Technology, 4800 Oak Grove Drive, Pasadena, CA 91109\label{NASAJPL}
        \and Department of Astronomy and Astrophysics, University of California, Santa Cruz, 1156 High Street, Santa Cruz, CA 95064 USA\label{UnivCalifornia}
        \and Department of Physics and Astronomy, University of California, Riverside, Riverside, CA 92521, USA \label{riverside}
        \and Zentrum f\"{u}r Astronomie, Universit\"{a}t Heidelberg, Philosophenweg 12, D-69120, Heidelberg, Germany \label{Heidelberg}
        \and LUX, Observatoire de Paris, Université PSL, Sorbonne Université, CNRS, 75014 Paris, France \label{MeudonLUX}
   }

   \date{Received; accepted}

\abstract{ 
While stellar mass correlates strongly with halo mass, it remains unclear whether halo mass alone governs galaxy star-formation activity, or whether secondary halo properties and environment also play a role. We investigate these effects beyond halo mass by measuring the auto- and cross-correlations of star-forming and quiescent galaxies in the COSMOS-Web survey from $z = 5$ to the present day. To isolate environmental contributions, we introduce a method that matches the halo mass distributions of both populations using the \textsc{UniverseMachine} model. We find that quiescent galaxies remain more strongly clustered than star-forming systems by at least $0.5-1$ dex at all redshifts, even after controlling for halo mass. At $z \le 2$, this excess clustering increases towards lower stellar masses, with the most clustered objects being $\log(M_\star/{\rm M}_\odot) \le 9.5$ quiescent galaxies. This points to environmental quenching significantly affecting low-mass galaxies at $z \le 2$, likely driven by ram-pressure stripping or the suppression of cold gas accretion, as these objects show disky morphologies. Cross-correlations further reveal one-halo conformity up to $z \simeq 2$: low-mass (or satellite) quiescent galaxies are more strongly clustered around massive (or central) quiescent galaxies than around star-forming centrals of the same halo mass. This signal may arise from quenching mechanisms affecting both centrals and satellites, correlated assembly histories prior to infall, or dependencies on secondary halo properties. Both environmental quenching and conformity appear to vanish between $z \simeq 5$ and $2$. Together, these results challenge the common assumption that clustering and star-formation activity depend solely on halo mass.
}


\maketitle

\section{Introduction} \label{sec:intro}

A tight relationship exists between halo mass and galaxy stellar mass assembly, as evidenced by measurements of the stellar-to-halo mass ratio (SHMR) over time \citep[\eg][]{White&Rees1978,Wechsler&Tinker2018}. This connection is therefore expected to play a role in the cessation of star formation, known as quenching. For instance, \cite{Dekel2006} suggests a critical halo mass $M_{\rm h} \simeq 10^{12}\,{\rm M}_\odot$ above which cold gas filaments cannot penetrate the halo any more because of gravitational shocks at the halo virial radius $-$ a mass scale which also corresponds to where active galactic nuclei (AGN) feedback is thought to quench galaxies \citep[\eg][]{Croton2006, Bower2006, Cattaneo2006}. Observations reveal that halos of this mass or greater host a higher proportion of passive galaxies, along with an increased fraction of satellite galaxies \citep{vandenBosch2003}. Such trends are empirically modelled as a function of halo mass by \cite{Peng2010}: it separates the ``mass quenching'', which primarily affects central galaxies through internal processes like AGN activity; versus the ``environmental quenching'', that mainly impacts satellites in massive halos (via ram-pressure stripping, gas depletion, etc.; \eg \citealt{Gunn&Gott1972, Dressler1980, Farouki&Shapiro1981, Peng2015}) or results from large-scale effects (differences in gas properties and accretion depending on the location in the cosmic web, \eg \citealt{Laigle2018, Song2021, Tojeiro&Kraljic2025}). 

Both effects are believed to drive the strong correlation of galaxy clustering (being the two-point correlation function of galaxy positions) with galaxy properties, such as colour or star formation rate (SFR). Redder galaxies are observed more clustered than their bluer counterparts \citep[\eg][]{Zehavi2005,Tinker2013,Berti2021}. This occurs primarily because quiescent galaxies (QGs) tend to reside in more massive halos and represent a larger fraction of satellites than star-forming galaxies (SFGs), enhancing their overall clustering. 

However, questions remain regarding environmental quenching and the evolution of its impact over time, especially since most statistical studies have been limited to $z < 3$. While correlations between star formation activity and various definitions of environment have been identified, some of these are contradictory, and establishing causality is challenging. For instance, some studies at $z \sim 1$ find that SFRs can be enhanced in locally overdense regions where gas can be more readily replenished \citep{Darvish2014}, while others suggest the opposite \citep{Peng2010, Wetzel2012}. Moreover, there is no consensus regarding the relative differences between the SHMR of QGs and SFGs \citep[\eg][]{More2011, RodriguezPuebla2015, Zu&Mandelbaum2016}. Isolating secondary environmental effects at fixed halo mass is complex, as measuring halo masses in observational data requires indirect modelling and strong assumptions \citep[\eg][]{Treyer2018}.

Additionally, it remains unclear whether secondary halo properties beyond just mass, such as halo assembly history, also influence star formation activity. If such effects exist, they might translate into the clustering of QGs and SFGs: this is what we call the ``galaxy assembly bias'' \citep[\eg][]{Croton2007,Zentner2014,Hearin2016b,Tinker2018a,Lyu2023}. Supporting evidence comes from \cite{Wang2022} and \cite{Zentner2019} for example, who argued that models assuming assembly bias based on halo concentration better reproduce their one and two-point galaxy statistics. According to \cite{Wang2026}, the gravitational potential of more concentrated halos may reduce gas outflows, leading to an enhanced chemical enrichment and stellar mass growth. In addition, the SFR of a galaxy may be correlated with that of its central, a phenomenon known as ``conformity'' \citep[\eg][]{Weinmann2006,Knobel2015,Kawinwanichakij2016,Hearin2016a,Berti2017}. For instance, \cite{Hatfield&Jarvis17} used auto- and cross-correlations up to $z = 2$ to demonstrate that lower-mass galaxies are more likely to be quenched when they are in proximity to a massive QG. Such processes may leave detectable imprints in galaxy clustering, and if significant, must be accounted for in models of the galaxy–halo connection.

Therefore, the key questions driving this work are: In which environments do QGs live, across cosmic time? And does halo mass alone suffice to explain galaxies' star-formation activity and spatial distribution? We address this by measuring auto- and cross-correlations of QGs and SFGs in the COSMOS-Web survey \citep{Casey23_CWeb}. A major strength of our approach is the use of wide and deep data from the James Webb Space Telescope (JWST), enabling the observation of QGs from $z = 0$ back to the earliest epochs at which they are observed \citep[$z \sim 5$; \eg][]{Carnall2023, Long2024, deGraaff2025}. This study builds on the previous work by \citealt{Paquereau2025} (hereafter \citetalias{Paquereau2025}) which investigated the SHMR in COSMOS-Web at $z = 0.1 -12$, and found a clear evolution of the galaxy-halo connection with redshift.

We begin in \refsec{sec:data} by describing the sample selection. After introducing the method in \refsec{sec:method}, clustering measurements are presented in \refsec{sec:results} for the auto-correlations of SFGs and QGs, and in \refsec{sec:crosscorr} for cross-correlations. We then discuss the implications for quenching and the galaxy-halo connection in \refsec{sec:discussion}, as well as the challenges of this study. Conclusions and future directions are presented in \refsec{sec:conclusion}.

Throughout this work, we adopt AB magnitudes \citep{OkeGunn1983_ABmag} and a cosmology from \cite{Planck18}. Stellar masses ($M_\star$) are computed assuming a \cite{Chabrier03} initial mass function. All logarithms in this paper are base 10.

\section{Data and sample selection} \label{sec:data}

\subsection{The COSMOS-Web survey and catalogues} \label{subsec:CWebsurvey}

As the largest contiguous extragalactic imaging survey conducted with JWST to date, COSMOS-Web is ideal to study the environment of galaxies across a wide range of scales and cosmic epochs. It benefits from deep high-resolution near-infrared imaging from JWST, covering an area of $0.54 \, {\rm deg}^2$ in the COSMOS field with a $5\sigma$ depth of $28.1 \, \mathrm{mag}$ in the F444W band. This is combined with the wealth of multi-wavelength data already available in COSMOS from 0.3 to 8 $\si{\micro\meter}$. From this dataset, a catalogue of over 700,000 galaxies with photometric measurements and properties has been created: COSMOS2025\footnote{This catalogue can be accessed at \url{https://cosmos2025.iap.fr/catalog.html}, along with the JWST mosaics and other data products as the group catalogue.} \citep{Shuntov2025d}. Photometric redshifts have been determined using the SED fitting code \LePHARE \citep{Arnouts02_LP, Ilbert06_LP}, achieving a great photo-$z$ accuracy characterised by a median absolute deviation $\sigma_{\rm MAD} \le 0.03$ and a median photo-$z$ uncertainty of $\Delta z/(1+z) \simeq  0.04$ across our redshift range. Non-parametric star formation histories and associated physical properties for all sources were derived with the code \CIGALE \citep{Boquien19_CIGALE, ArangoToro2024}. For a more detailed overview of COSMOS2025, one can refer to \citetalias{Paquereau2025} and \cite{Shuntov2025d}.

Additionally, \cite{Toni2025} built a group catalogue up to $z \simeq 3.7$ in COSMOS-Web. Groups are identified using the algorithm \texttt{AMICO} \citep[][]{Bellagamba2018}. It is a linear optimal matched filter which models galaxy density as a combination of a ``signal'' component $-$ representing the contribution of group galaxies (modelled using a \citealt{NFW1997} profile for their distribution within halos, and a \citealt{Schechter1976} function for their number density) $-$ and a ``noise'' component from field galaxies (modelled using the distribution of the whole sample). The algorithm seeks to determine where the signal-to-noise ratio S/N of the galaxy density is maximised. In total, 1678 candidate groups have been identified with a S/N $> 6$ and purity $P > 77\%$, of which half are confirmed with spectroscopy or are matched with known overdensities.

In this work, we adopt the same sample cleaning procedure as in \citetalias{Paquereau2025}, excluding stars, quasars, sources within star masks, objects fainter than $m_{\textrm{F444W}} = 27.75$ mag, and galaxies with poorly constrained redshift probability distribution functions.

\subsubsection{Star-forming/quiescent classification}  \label{subsec:classification}

The first step of this work is to establish a classification separating galaxies actively forming stars (SFGs) and the ones that do not (QGs), which remains consistent across our $z = 0 - 5$ range. This is a challenging task, because as redshift increases, the SFG/QG bimodality becomes weaker, QGs are rarer, or may exhibit a wider variety of properties than in the local Universe \citep[\eg][]{Valentino2020}. 

We adopt an approach that sets an evolving threshold for the galaxies' sSFR (${\rm sSFR} = {\rm SFR}/M_\star$), as introduced by \cite{Fontana2009}. It sets a certain timescale for star formation relative to the age of the Universe at the galaxy's redshift: ${\rm sSFR} \le \beta \, / \, t_{\rm H}(z)$ to select QGs, where $t_\mathrm{H}(z)$ is the Hubble time. Although $\beta$ is arbitrary, most works use $\beta = 0.2$ \citep[\eg][]{Pacifici2016,Carnall2018,Tacchella2019}. It is broadly equivalent to a $UVJ$ selection and a threshold of $\log({\rm sSFR}/{\rm yr}^{-1}) \simeq -10.9$ at $z < 1$, but decreases to $-10.15$ at $z = 3.5$. We also use $\beta = 0.2$ since it appears to best delineate our $z - \mathrm{sSFR}$ plane, and aligns with the gradual increase of the normalisation of the main sequence (MS) with $z$. 

Although the sSFR classification depends on the SED model and may be influenced by uncertainties in redshift and physical parameter estimates, it utilises most of the available data and can be consistently applied across all redshift bins as well as in other surveys or simulations with the same prescription (which is mandatory in our case, in order to use the \textsc{UniverseMachine} model; see \refsubsec{subsec:massmatching}). A comparison with alternative classifications is discussed in \refapp{app:class}.

\subsection{Completeness and bining}  \label{subsec:bins}

\begin{figure}[h!]
    \centering
    \includegraphics[width=1.0\columnwidth]{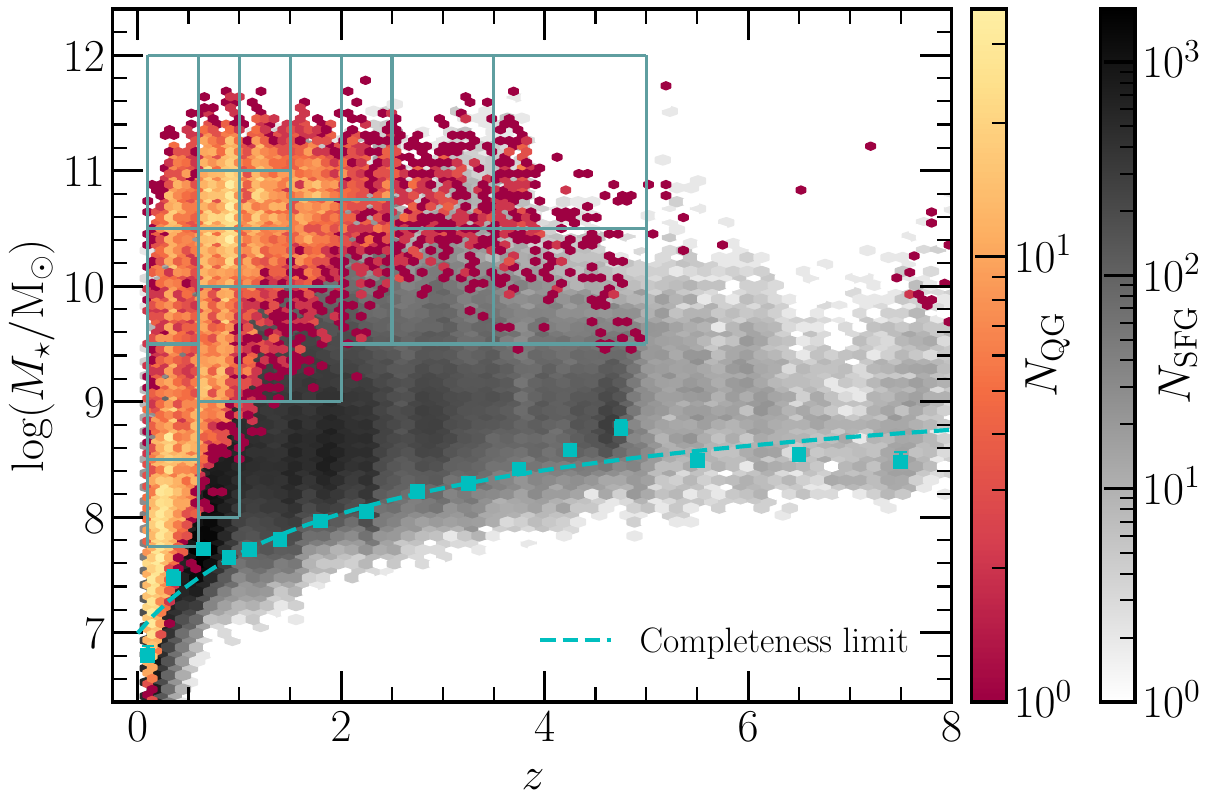}
    \vspace{-6mm}
    \caption{Stellar mass completeness in the $z - M_\star$ diagram. Hexagons represent counts of QGs (with a red gradient) and SFGs (grey gradient). The redshift bins used to measure clustering with mass thresholds are indicated in the foreground.}
    \label{fig:completeness}
\end{figure}

Secondly, we construct stellar mass-complete subsamples to avoid selection biases that could distort the clustering signal. We use \cite{Pozzetti10} formula to get the mass completeness limit for the whole sample\footnote{The completeness limit for the whole sample is identical to that of SFGs alone, as the faintest objects of the sample are exclusively star-forming; while QGs, brighter because of their higher mass-to-light ratios, lie well above the survey magnitude limit.} (see Sect.~2.2.1 in \citetalias{Paquereau2025}). Redshift and stellar mass bins above the completeness limit are chosen to contain at least $\sim 100$ objects per bin and roughly the same order of magnitude for the numbers of SFGs and QGs. Redshift bin widths correspond to approximately similar cosmic timescales of $0.7-1.5$\,Gyr (except at $z < 1$). \reffigure{fig:completeness} presents the sample completeness, along with the bins used in this work. We also show numbers and quiescent fractions as a function of mass and redshift in \refapp{app:numbers} and \ref{app:fQG}. In total, the sample comprises 76506 SFGs and 8750 QGs.

\subsection{Halo-mass matching} \label{subsec:massmatching}

One of the objectives of this study is to investigate the effects beyond halo mass on galaxy clustering. Thus, we want to construct SFG and QG samples that share a similar halo mass distribution, so that any remaining clustering differences must arise from secondary effects. This is, however, a complex task because we lack direct information on this quantity. Not robustly controlling for halo mass can lead to misleading interpretations of conformity or assembly bias \citep[\eg][]{Calderon2018}.

To do so, we have recourse to the semi-empirical model \UniverseMachine (hereafter \UM; \citealt{Behroozi19}), built in an iterative way to reproduce various $z = 0-10$ observations (stellar mass or UV luminosity functions, clustering, etc.), including from the COSMOS field. From the modelled \UM COSMOS lightcone, we selected SFGs and QGs the same way as described in \refsubsec{subsec:classification}. We confirmed that just matching the stellar mass distributions of SFGs with QGs is far from matching their underlying halo mass distributions: for instance, at fixed $M_\star \ge 10^9 \,{\rm M}_\odot$, the median halo mass of QGs is 10 times higher than that of SFGs. Instead, we developed a ``halo-mass matching'' (HMM) method that makes use of \UM to coincide QGs and SFGs halo mass distributions, following these steps:
\begin{enumerate}
  \item Match the shape of stellar mass distributions of \UM to COSMOS-Web, for both SFGs and QGs. This ensures that we have comparable populations in both the model and observations. 
  \item In \UM, match the halo mass distributions between SFGs and QGs (previously stellar mass-matched), in terms of shape and amplitude. We return \UM's SFGs/QGs new stellar mass distributions after this matching for the next step. 
  \item Modify the COSMOS-Web SFGs/QGs initial stellar mass distributions to match the shape of \UM's halo mass-matched SFGs and QGs stellar mass distributions.
\end{enumerate}
In each of these steps, distributions are matched by randomly excluding sources from the samples. We repeat this HMM procedure 20 times, measuring clustering independently each time. The final signal is taken as the median over all iterations, retaining only data points with ${\rm S/N} > 1$, and associated uncertainties are given by the quadrature sum of the median clustering error and the standard deviation across iterations. This ensures stable results: the scatter across iterations remains at most 20\% of the median clustering uncertainty for SFGs and 40\% for QGs. An illustration of this process can be found in \refapp{app:HMM}.

\section{Method} \label{sec:method}

We measure angular two-point correlation functions (ACF) $w(\theta)$ of galaxy positions separated by an angle $\theta$ using the formula from \cite{LandySzalay93}. For a random distribution $R$ and two populations of number counts $D_1$ and $D_2$, the cross-correlation between the two in the bin $[\theta, \theta + \delta \theta]$ is given by:
\begin{equation}\label{eq_ch7:crossw}
    w(\theta) = \frac{D_1 D_2 - D_1 R - D_2 R + RR}{RR} \; .
\end{equation}
This becomes an auto-correlation function if $D_1 = D_2$. Following \citetalias{Paquereau2025}, we construct the random catalogue on the same area as COSMOS-Web, accounting for the same masks, and containing about $50$ times the total number of objects in our sample.

We measure these correlations with the algorithm \texttt{treecorr} \citep{Jarvis_treecorr}, and errors are computed using the jackknife resampling method. It divides the area into patches ($N_{\rm patches} = 20$ here) and calculates the covariance matrix of the ACF by removing one patch at a time. An additional source of errors comes from the finite survey area, which can lead to an underestimation of the clustering at large scales: this is the integral constraint \citep{Groth&Peebles1977}. Its impact becomes significant at scales above $0.02\,\mathrm{deg}$. A correction term exists \citep{RocheEales99}, but as it necessitates an estimation of the ``true'' ACF, we do not apply it in our analysis. We calculate the ACF across a range of angular scales determined to be above the resolution and below the total area of the survey: $10^{-4} \le \theta \le 10^{-1} \,\mathrm{deg}$, with a number of bins $N_\theta \in [10, 15]$ (depending on $z$). 

\section{Auto-correlation functions of star-forming and quiescent galaxies} \label{sec:results}

First, we present measurements of auto-correlation functions of mass-complete subsamples of SFGs and QGs across $z = 0.1 - 5$ in COSMOS-Web.

\subsection{Samples limited by mass thresholds} \label{subsec:autocorr_massthresh}

\begin{figure*}[p!]
    \centering
    \includegraphics[height=0.95\textheight]{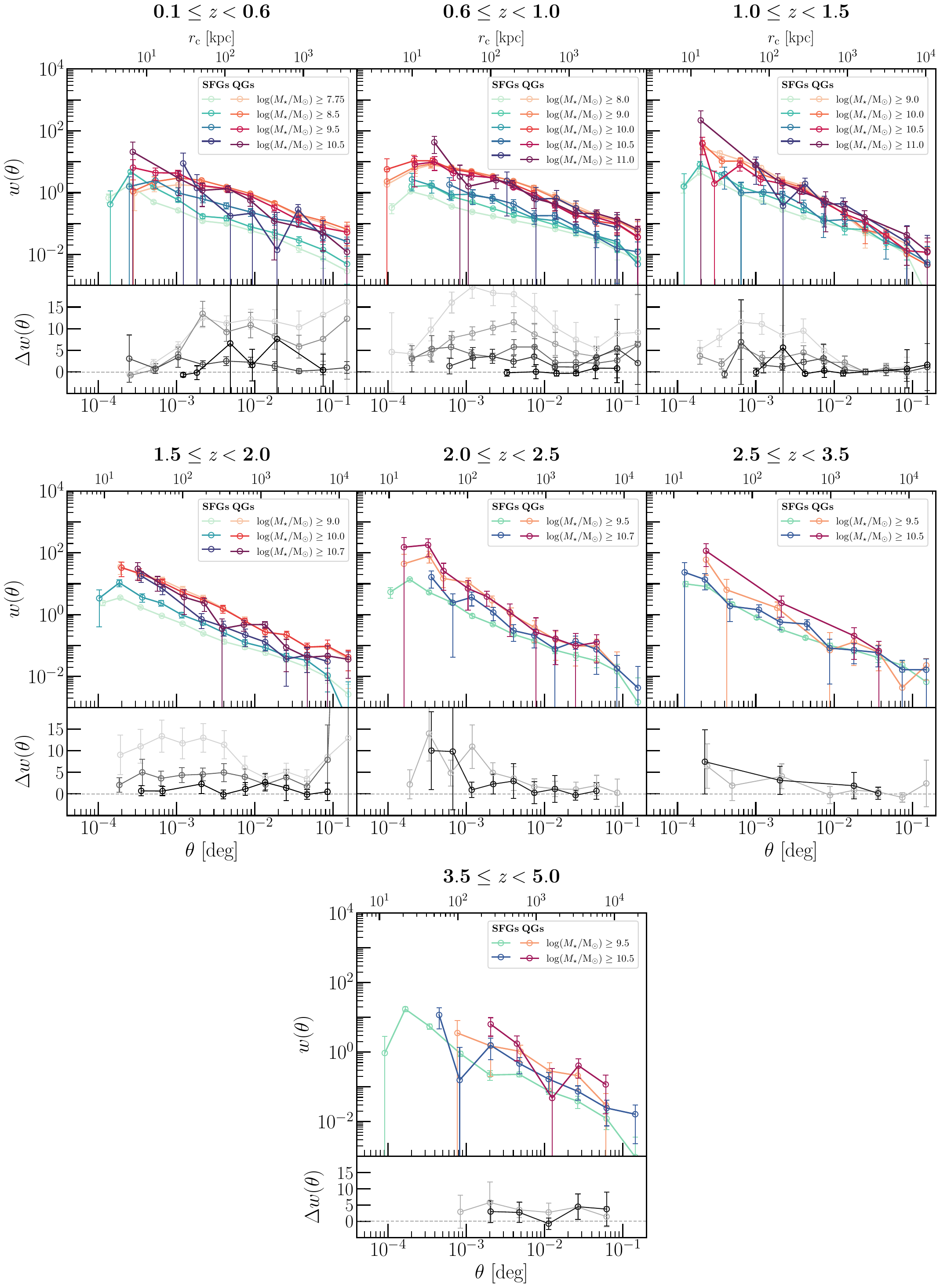}
    \caption{Angular auto-correlation function for quiescent (red lines) and star-forming (blue lines) galaxies, in samples limited by lower limits in stellar mass and redshift bins. Both angular ($\theta$) and comoving physical scales ($r_{\rm c}$) are represented, where the latter has been calculated at the bin mean redshift. The relative difference $\Delta w$ between the clustering of SFGs ($w_{\rm SFG}$) and QGs ($w_{\rm QG}$) is shown in each bottom panel for the same mass limits (with a grey scale, where lighter grey corresponds to a lower mass threshold). This is done without halo mass matching.}
    \label{fig:autocorr_thresh}
\end{figure*}

The auto-correlation is shown for samples limited in stellar mass, first without HMM in \reffig{fig:autocorr_thresh}. 
We find that QGs are more clustered than SFGs of about 0.5 to 1.2\,dex \citep[an common result in the literature, \eg][]{Zehavi2011,RodriguezPuebla2015,Coil2017,Berti2019}, consistently across all redshifts. The only exception to this pattern occurs for galaxies with $\log(M_\star/{\rm M}_\odot) \ge 11$, where the clustering of QGs and SFGs is similar (a result that holds despite comparable numbers of QGs and SFGs in this bin). The relative ratio $\Delta w = (w_{\rm QG}-w_{\rm SFG})/w_{\rm SFG}$ becomes less significant with increasing redshift, decreasing from $\ge 10$ at $z < 1$ to $\Delta w \le 5$ at $z \ge 3.5$ for the lower mass bin. In other words, the environment of low-mass QGs is more similar to that of SFGs at higher $z$, but we still find that there are differences in their environments even at $3.5 \le z < 5$. Another interesting finding is that $\Delta w$ anti-correlates with the stellar mass threshold, particularly at $z < 2$ and for the intra-halo term ($\theta \lesssim 10^{-2}\,{\rm deg}$ at $z \simeq 1$, or $r < 1\,{\rm Mpc}$). The clustering of lower mass QGs is then stronger than that of higher mass QGs, while the opposite trend is observed for SFGs. 

After conducting halo mass matching, some of the ACFs become poorly constrained because we significantly reduced the number of objects in each bin (by about 20 to 40\% for QGs). This is presented in \refapp{app:autocorr_thresh_HMM}. Nonetheless, the trends still show that QGs exhibit a clustering 0.2 to $1 \,\mathrm{dex}$ higher compared to SFGs in the range $z = 0.1 - 5$. However, the variations in clustering for QGs based on stellar mass are diminished, as measurements in all mass bins for QGs are consistent with each other within the uncertainties.

\subsection{Samples in mass bins} \label{subsec:autocorr_massbins}

\begin{figure*}[b!]
    \centering
    \includegraphics[width=0.95\textwidth]{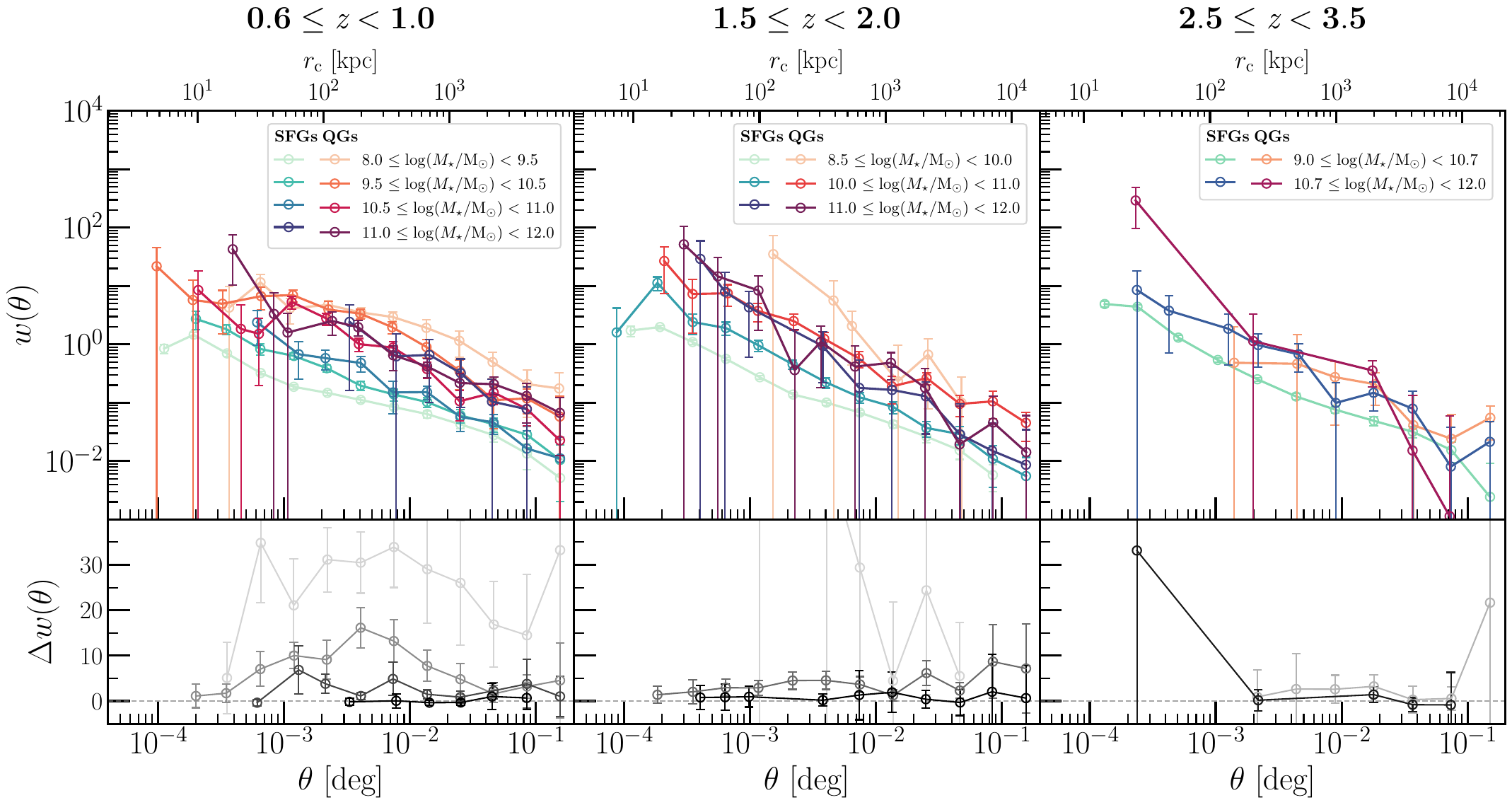}
    \caption{Similar to \reffig{fig:autocorr_thresh}, but considering stellar mass ranges instead of thresholds, in three redshift bins. This is done without halo mass matching.}
    \label{fig:autocorr_ranges}
\end{figure*}

To better separate the contributions from different populations, we also compute the clustering of SFGs and QGs within stellar mass ranges instead of thresholds. This is shown for three redshift bins in \reffig{fig:autocorr_ranges} without HMM, and in \refapp{app:autocorr_thresh_HMM} with HMM. 
The signal is however noisier as the number of QGs in each bin is significantly reduced. Because of this lack of statistics, we have not been able to perform HMM at $z \ge 2$. Nevertheless, this allows us to confirm that QGs become more clustered as stellar mass decreases at $z < 2$, which contrasts with the behaviour of SFGs. For instance, up to $z = 2$, QGs with $M_\star < 10^{10}\,{\rm M}_\odot$ are about 20 times more clustered than SFGs of same mass, and across all angular scales.

While these results are supported by other literature measurements (\eg \citealt{McCracken2008}, finding faint red galaxies more clustered than their brighter counterparts; or \citealt{Tinker2013}; \citealt{Sato2014}; \citealt{Ji18}), it is in contradiction with the hierarchical model of structure formation. Indeed, this scenario would place more massive galaxies in more massive halos, leading to greater clustering with increasing stellar mass. This is the case for SFGs, but the contradiction for QGs may be attributed to how they populate halos or to the quenching mechanisms at work in high-density environments (these low-mass QGs could then be satellites, affected by environmental quenching). We discuss this in more detail in \refsubsec{subsec:envquenching}. 

\subsection{Synthesis of auto-correlations} \label{subsec:autocorr_summary}

\begin{figure}[ht!]
    \centering
    \includegraphics[width=1.0\columnwidth]{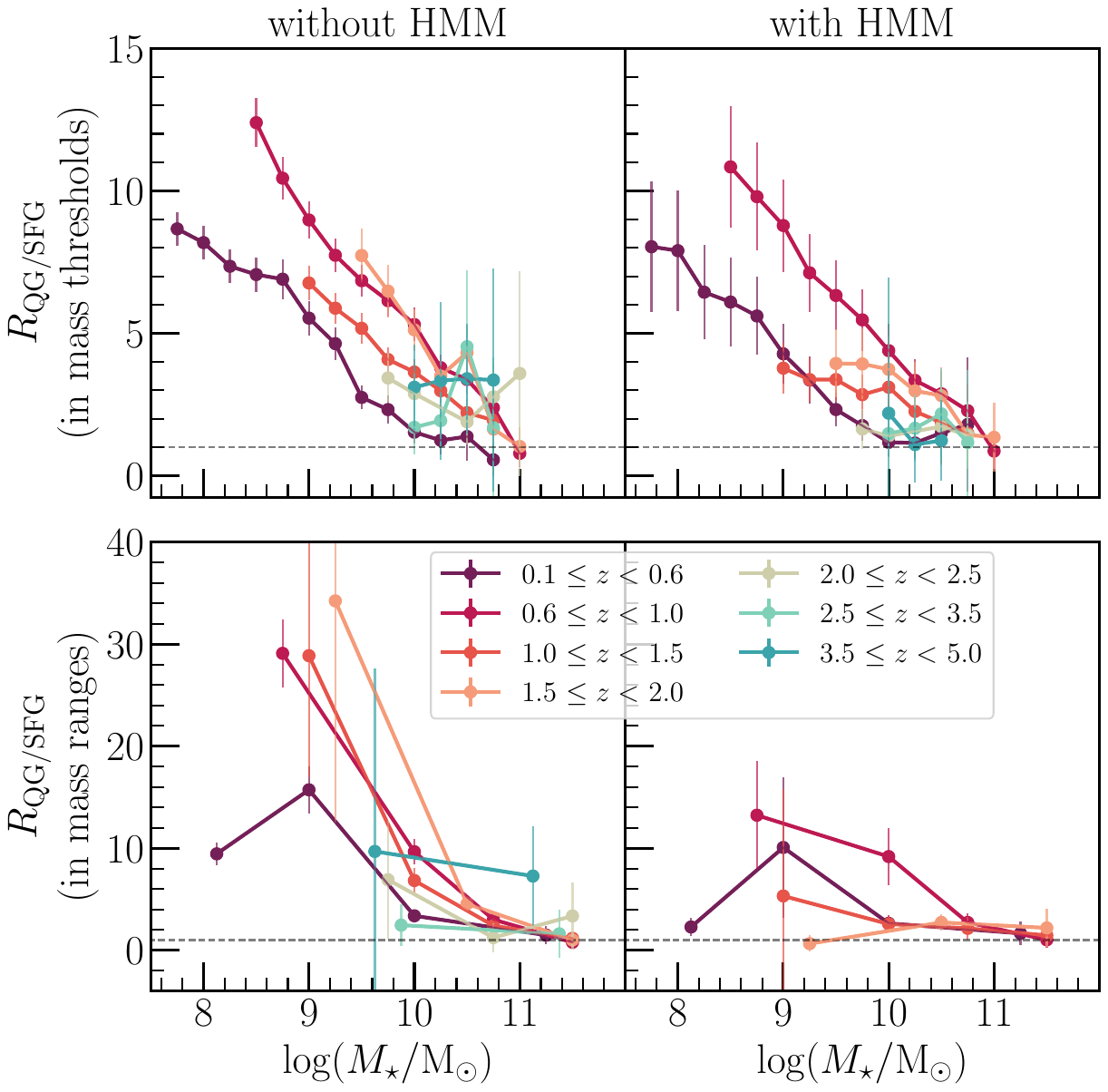}
    \vspace{-4mm}
    \caption{Ratio of clustering amplitudes $R_{\rm QG/SFG} = A_{\rm QG}/A_{\rm SFG}$ as a function of stellar mass and redshift, for one-halo spatial scales ($r_{\rm c} < 1 \,{\rm Mpc}$). Top row: for stellar mass thresholds (\refsubsec{subsec:autocorr_massthresh}, with more thresholds), with and without HMM. Bottom row: for stellar mass ranges (\refsubsec{subsec:autocorr_massbins}), with and without HMM. A line at $R_{\rm QG/SFG} = 1$ is shown in dashed grey.}
    \label{fig:autocorr_sum}
\end{figure}

We synthesise our results in \reffig{fig:autocorr_sum} by representing the ratio of the QG/SFG clustering $R_{\rm QG/SFG}$ as a function of $M_\star$, for the different cases considered.
To average over angular scales and account for statistical variations, we model each ACF as two simple power laws $w^{\rm mod}(\theta) = A \theta^{-\gamma}$ in the one-halo and two-halo regimes ($r_{\rm c}$ lower or greater than $1 \,{\rm Mpc}$, respectively). Each one-halo and two-halo ACF is fitted while assuming a shared $\gamma$ for QGs and SFGs\footnote{
While the literature shows that QGs tend to have steeper ACF slopes than SFGs, the difference remains below $\sim$20\% in our angular scale regime \citep[\eg][]{Sato2014,Coil2017}, which would not significantly affect our conclusions.
} in each bin, as we are only interested in the amplitude ratio. The ratio of fitted amplitudes $R_{\rm QG/SFG} = w_{\rm QG}^{\rm mod}/w_{\rm SFG}^{\rm mod} = A_{\rm QG}/A_{\rm SFG}$ is then computed, along with its associated fitting errors.

As shown, $R_{\rm QG/SFG}$ is significantly higher for low-mass galaxies, with low-mass QGs being globally 5 to 13 times more clustered than SFGs with mass thresholds, or 10 to 30 times more than SFGs in mass ranges below $\sim 10^{9.5}\,{\rm M}_\odot$. This ratio decreases gradually with increasing $M_\star$ (for both with and without HMM), except for the lowest mass bin at $z \le 0.6$ (see \refsubsec{subsec:envquenching}). This trend persists up to $z = 2$; however, beyond this redshift, the ratio appears to remain constant with $M_\star$ $-$ an observation that becomes even clearer after applying HMM. We interpret these trends in \refsubsec{subsec:envquenching}.

\section{Cross-correlations and search for conformity} \label{sec:crosscorr}

Our first results suggest that environmental effects quench low-mass galaxies at $z \lesssim 2$, and leave a signature in the clustering of QGs. We now ask whether the star formation activity is correlated among galaxies sharing a common halo or large-scale environment, beyond halo mass. To explore this possibility, we turn to cross-correlations of SFGs and QGs subdivided into two classes chosen to trace different environments. 

\subsection{Cross-correlations with high- and low-mass samples} \label{subsec:crosscorr_mass}

Ideally, we would separate galaxies into centrals and satellites, since centrals are thought to trace halo mass and the large-scale structure, while satellites are more influenced by the halo environment. However, accurately identifying centrals and satellites is complex and may introduce biases \citep[\eg][]{Sun2018, Tinker2018a, Calderon2018}. Therefore, we start by splitting our sample into high- and low-mass galaxies following the method proposed by \cite{Hatfield&Jarvis17}, which assumes that high-mass galaxies act as ``drivers'' of environmental processes, while the low-mass ones are the ``followers'' affected by these effects. We aim to extend their analysis (limited to $0.4 \le z < 2$) to a deeper dataset thanks to COSMOS-Web. 

We define the high- and low-mass classes to roughly align with the mass distributions of the central and satellite populations (respectively), while also ensuring sufficient statistics in each bin. According to the study of \cite{Zaidi2024} done in COSMOS, a threshold of about $\log(M_\star / {\rm M}_\odot) \ge 10.5 $ corresponds to an observed decline of the satellites' mass function, beyond which centrals largely dominate the counts. Consistently, the COSMOS-Web group catalogue shows that $\sim$90\% of satellites fall below this threshold, while $\sim$50\% of centrals (and $\sim$90\% of quiescent centrals) lie above it. Hence, we choose the low-mass class for both SFGs and QGs, as follows: $8.5 \le \log(M_\star/{\rm M}_\odot) < 10$ for $z < 1.5$, $9 \le \log(M_\star/{\rm M}_\odot) < 10.5$ for $1.5 \le z < 2.5$ and $9 \le \log(M_\star/{\rm M}_\odot) < 10.75$ for $z \ge 2.5$. High-mass subsamples consist of objects with masses above these upper limits. 

\begin{figure*}[!htbp]
    \centering
    \vspace{1cm}
    \includegraphics[width=1.0\textwidth]{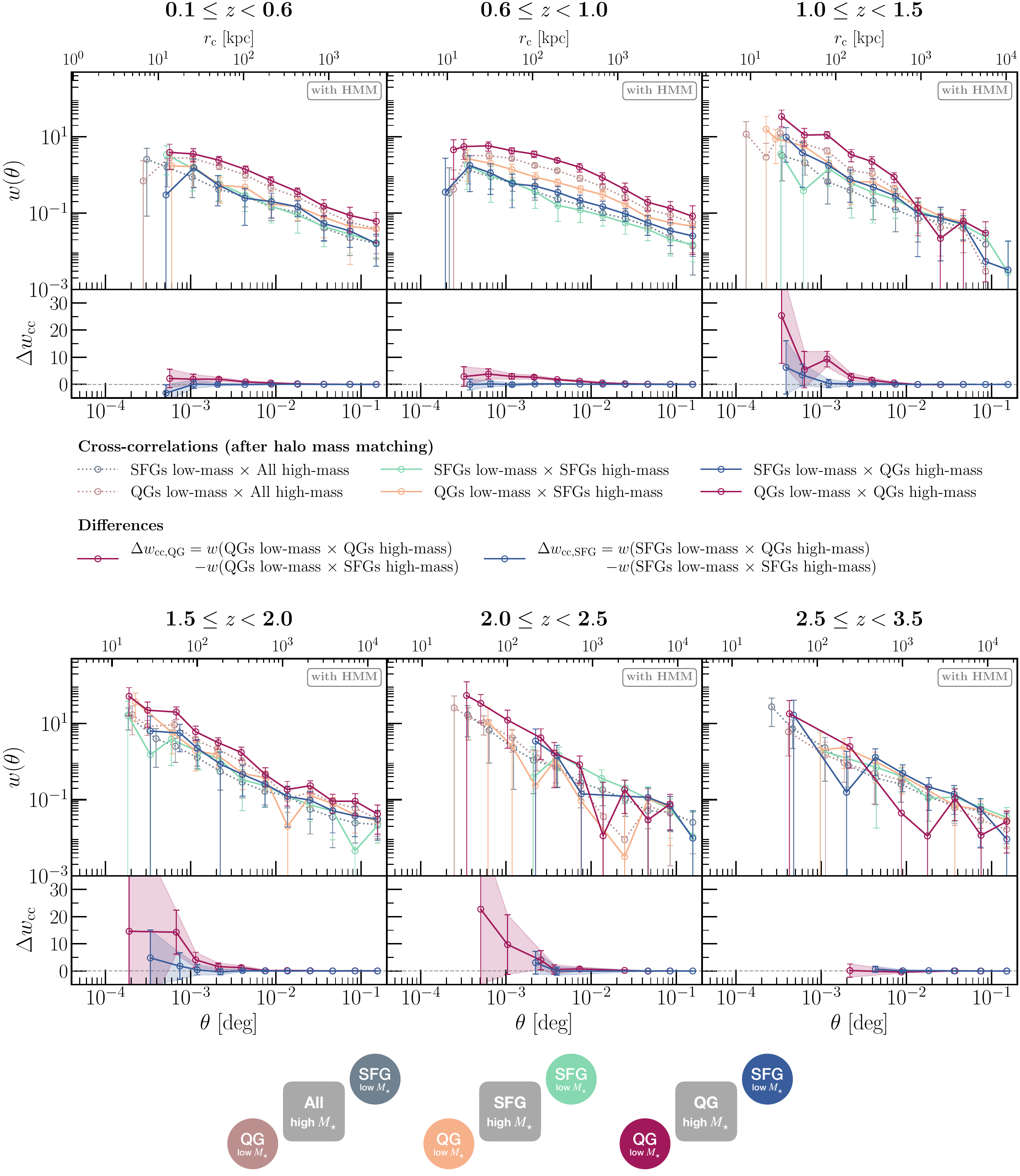}
    \caption{Angular cross-correlations for SFGs and QGs in redshift bins, split in high- and low-mass subsamples, and after performing halo mass-matching. Differences between cross-correlations $\Delta w_{\rm cc}$ involving low-mass QGs and low-mass SFGs are shown in the bottom panels.}
    \label{fig:crosscorr_SM_HMM}
\end{figure*}

Controlling for halo mass is also crucial here, so we show cross-correlation measurements with HMM in \reffig{fig:crosscorr_SM_HMM} (and without in \refapp{app:crosscorr_withoutHMM}). One interesting result is that, at scales $\theta \lesssim 10^{-2} \,{\rm deg}$, the cross-correlation [QGs low-mass $\times$ QGs high-mass] is higher than [QGs low-mass $\times$ SFGs high-mass]. Specifically, the difference is significant
\footnote{Here, the detection significance $S$ is the S/N of the cross-correlation difference $\Delta w_{\rm cc}(\theta)$ integrated over $\theta$ (in one and two-halo regimes), relative to the null hypothesis $\Delta w_{\rm cc}=0$ (no conformity). We compute it via $S = {\rm PPF}(1 - p/2)$ where PPF is the inverse normal cumulative density function and $p$ is the $p$-value of a $\chi^2 = \sum_{\theta_i} (\Delta w_{\rm cc}(\theta_i)/\sigma_{\Delta w_{\rm cc}})^2$ distribution.} 
at a $\ge 3\sigma$ level at $z \le 1.5$ and becomes minor ($\le 1\sigma$) at $z \ge 2$ where the uncertainties make it also consistent with zero. It lies below $2\sigma$ at large scales. On the contrary, minor differences (below $0.5\sigma$) are seen between cross correlations [SFGs low-mass $\times$ SFGs high-mass] and [SFGs low-mass $\times$ QGs high-mass].

This observation suggests the detection of a one-halo conformity signal: quiescent low-mass galaxies are more likely to be found near quiescent massive galaxies than star-forming ones, even after accounting for halo mass, while the spatial distribution of star-forming low-mass galaxies does not depend on the type of the massive galaxy sample. However, this effect is observed only at the scale of a halo ($\lesssim 1 \,\mathrm{Mpc}$) and at $z \le 2$.

\subsection{Cross-correlations with centrals and satellites} \label{subsec:crosscorr_group}

Secondly, we make use of the COSMOS-Web group catalogue at $z \le  3.7$ to compute the cross-correlations with central and satellite populations in this range. Centrals are identified by \citealt{Toni2026}, considering these criteria: (1) proximity to the group centre in projected sky coordinates (distance $< 0.3\times R_{200}$, where $R_{200}$ is the group model radius used by the detection algorithm) and in redshift ($\Delta z < 0.05 \times [1+z]$); (2) having the highest stellar mass; and (3) a higher group membership probability if the first and second more massive galaxies have a mass difference lower than $0.25$. Satellites are then identified as group galaxies (having $> 50\%$ probability of belonging to a group) that are not centrals. The rest are labeled as ``field'' galaxies; these are not used in our work.

\begin{figure*}[!htbp]
    \centering
    \includegraphics[width=1.0\textwidth]{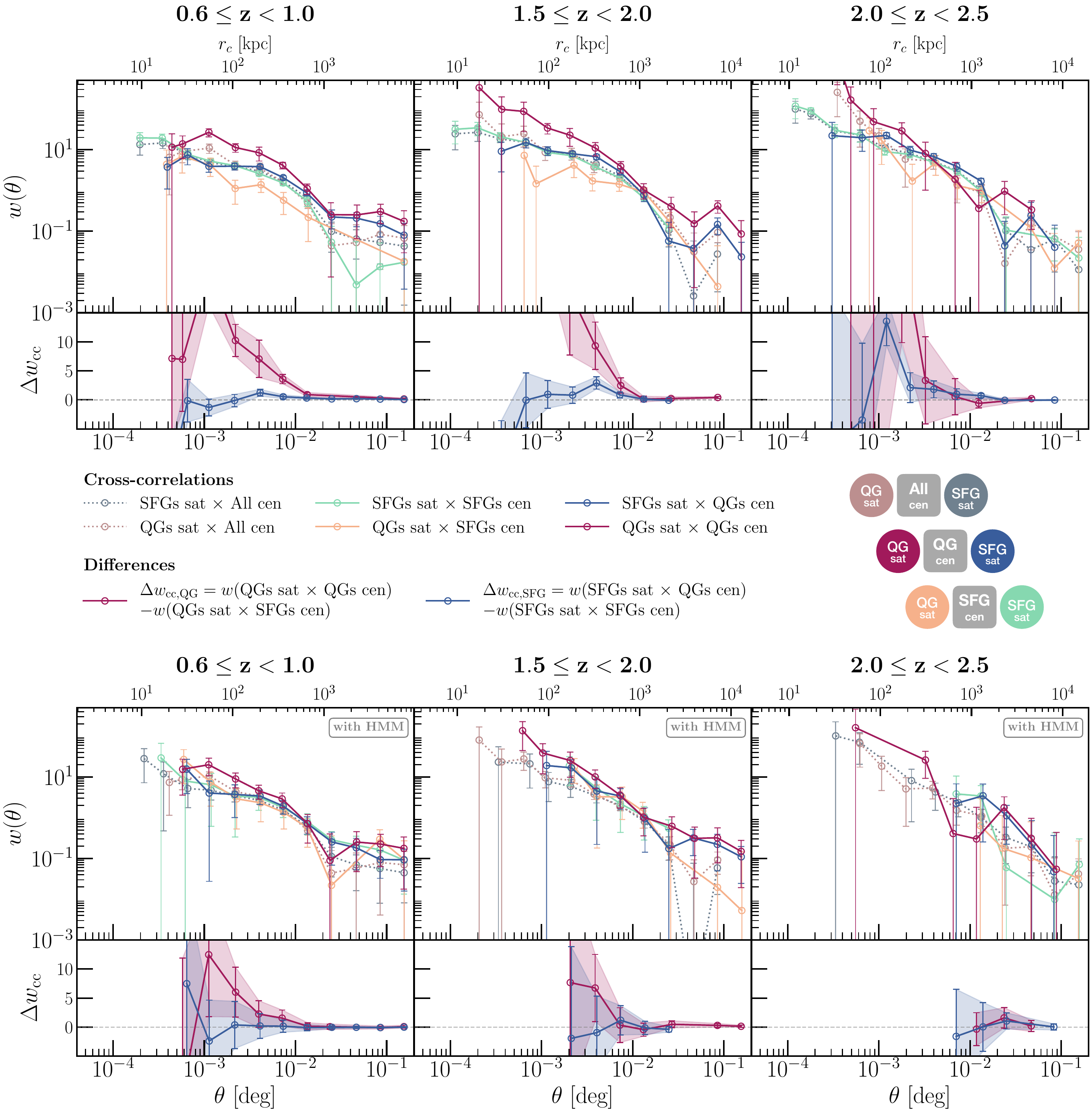}
    \vspace{-3mm}
    \caption{Angular cross-correlations for SFGs and QGs, subdivided in samples of satellites and centrals, in three redshift bins. Differences between samples involving quiescent satellites and star-forming satellites are shown in the bottom panels. Top row: without halo mass-matching. Bottom row: with halo mass matching.}
    \label{fig:crosscorr_group}
\end{figure*}

The results are shown in \reffig{fig:crosscorr_group} for three redshift bins, with and without halo mass matching. Cross-correlations without HMM are in agreement with our previous findings, with a conformity signal for quiescent satellites around quiescent centrals at $6\sigma$ for the bin $0.6 \le z < 1$, $> 1\sigma$ at $1 \le z < 1.5$, $3\sigma$ at $1.5 \le z < 2$ and becomes insignificant at $z \ge 2$. This is however largely reduced by matching halo mass distributions, with $\sim 1\sigma$ detections only for the bin $0.6 \le z < 1$ (although the signal remains visually apparent in $z < 2$ bins in \reffig{fig:crosscorr_SM_HMM}).

\subsection{Synthesis of cross-correlations} \label{subsec:crosscorr_summary}

\begin{figure}[t!]
    \centering
    \includegraphics[width=1.0\columnwidth]{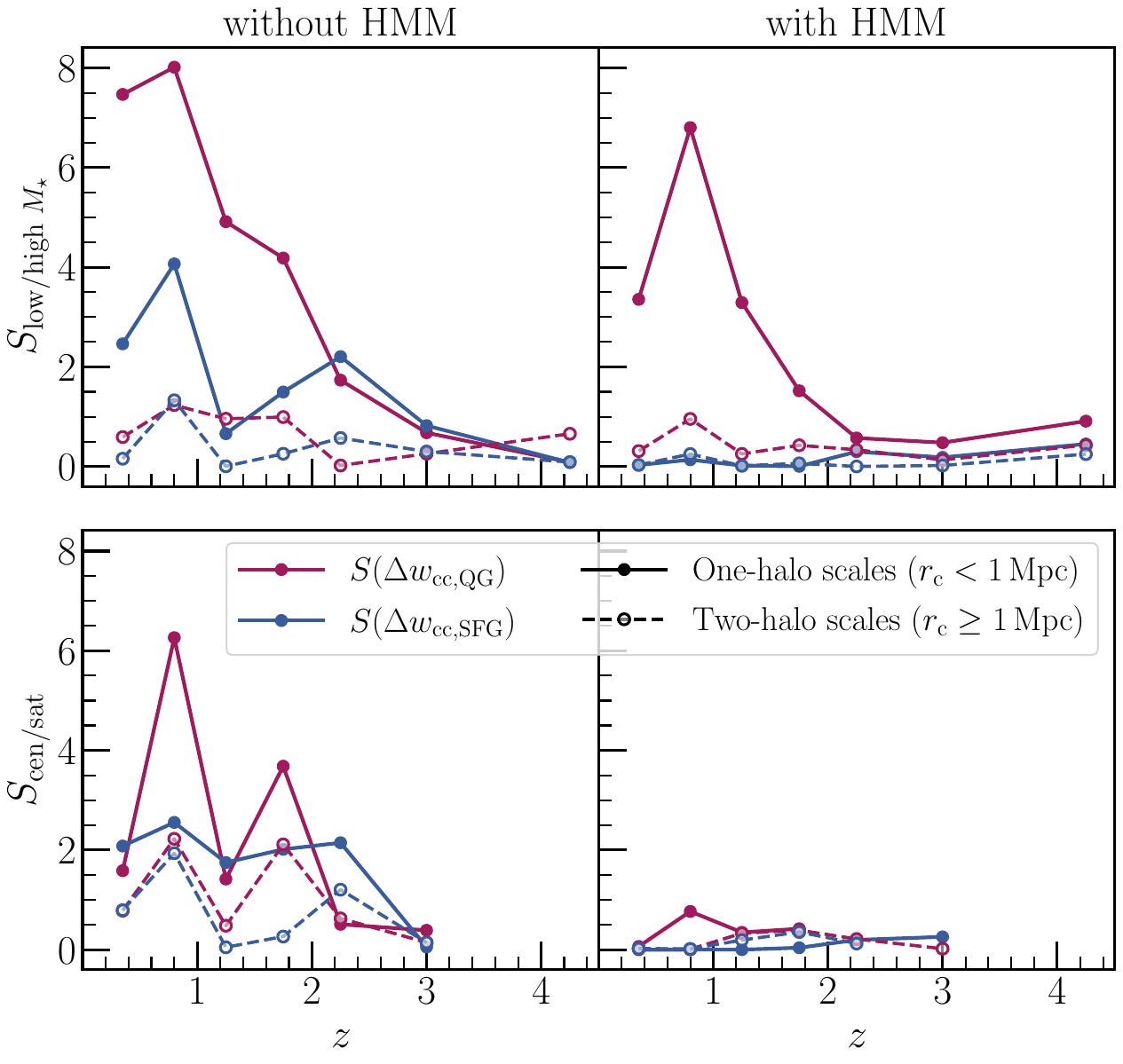}
    \vspace{-5mm}
    \caption{Detection significance of the conformity signal $S$ for QGs (red) and SFGs (blue) as a function of redshift, in the one-halo (solid lines) and two-halo (dashed lines) regimes. Top row: from low/high mass cross-correlations (\refsubsec{subsec:crosscorr_mass}), with and without HMM. Bottom row: from central/satellite cross-correlations (\refsubsec{subsec:crosscorr_group}), with and without HMM.}
    \label{fig:crosscorr_sum}
\end{figure}

These results are summarised in \reffig{fig:crosscorr_sum}, which shows the detection significance of conformity (denoted $S$) as a function of redshift, for low-mass (or satellite) QGs and SFGs around their central (measured from $\Delta w_{\rm cc, QG}$ and $\Delta w_{\rm cc, SFG}$ in \reffig{fig:crosscorr_SM_HMM} and \reffig{fig:crosscorr_group}). It covers the various cases considered (with/without HMM), and separates the one-halo and two-halo scales. This highlights that conformity is detected only in the one-halo regime, where low-mass QGs are more clustered around high-mass QGs than SFGs of the same halo mass. This observation appears at $z \le 2$ and becomes more pronounced as $z$ decreases. This is also observed for satellite QGs around central QGs, although the significance diminishes considerably after matching halo mass distributions. We caution that the two-halo term is subject to the integral constraint, which limits conclusions in this regime. These results are discussed in \refsubsec{subsec:conformity}.

\section{Discussion: what factors beyond halo mass influence galaxy clustering?} \label{sec:discussion}

\subsection{Environmental quenching up to $z \simeq 2$}   \label{subsec:envquenching} 

One of the key findings from \refsec{sec:results} is that QGs are always more clustered than SFGs of the same stellar mass. This indicates that even the earliest passive galaxies at $z \sim 3.5-5$ live in different environments compared to their star-forming counterparts. Applying HMM reduced this excess, particularly at $z \ge 2$, suggesting that the tendency of QGs to occupy more massive halos dominates over secondary effects. 
Yet, differences remain and may be attributed to secondary factors: QGs may predominantly be satellites, be connected to halo secondary properties, or occupy higher-density regions. We believe the latter alone cannot account for the $z > 1$ signal, since some studies show that the anti-correlation between sSFR and local density breaks down at $z \leq 1$, either by reversing to positive \citep[\eg in COSMOS,][]{Shi2024, Taamoli2024, Li2025} or becoming restricted to extreme overdensities only \citep{Hatamnia2026}. Instead, the galaxy's location within the cosmic web might regulate its star formation at fixed halo mass, as supported by \cite{Jego2026} who find a modulation of sSFR with distance to filaments in COSMOS at $z = 0.5-2$.

We note that there is an exception at $\log(M_\star/{\rm M}_\odot) \ge 11$, where QGs and SFGs clustering converge. Such massive SFGs are often central members of massive clusters, and are likely to become quenched due to gas exhaustion or the high probability of hosting an AGN at this mass. It is thus reasonable that they exist in environments similar to those of massive QGs. 

A second exception occurs for $\log(M_\star/{\rm M}_\odot) < 9$ galaxies at $0.1 \le z < 0.6$, showing a more comparable clustering between QGs and SFGs than at higher mass. \cite{Geha2012} indicate that quenched dwarfs are exclusively satellites, which are at a more advanced stage of environmental quenching than the star-forming dwarfs in the same halo; but further investigation of dwarf quenching is needed to explain this result.

The second key result from our measured auto-correlations is that as stellar mass increases, SFGs become more clustered $-$ which aligns with the hierarchical picture where more massive galaxies reside within more massive halos, themselves more clustered $-$ while QGs show the opposite trend. It persists even after HMM, but only up to $z \simeq 2$. We also observe the same trend in \UM by computing auto-correlations in halo mass bins instead of stellar mass (see \refapp{app:autocorr_UM}): at the same $M_{\rm h}$, QGs are more clustered than SFGs, and differences are greater as $\Mh$ decreases. This result is consistent with environmental quenching of low-mass galaxies: the highly clustered low-mass QGs are likely to be either satellites in groups or clusters, or galaxies influenced by their local environment \citep[\eg][]{Boselli2006, Moutard2018, Alberts2024}. This signal is present from $z \simeq 2$ to the present, consistent with other works finding no evidence for dominant environmental quenching above $z \sim 3$ \citep[\eg][]{Singh2025,Shuntov2026b}. 
This does not imply that environmental quenching does not exist at higher $z$; it may still operate across all masses to a lesser extent, and/or only in extreme overdensities such as protoclusters. Our remaining excess clustering for QGs at $z > 2$ after HMM could therefore reflect either of these effects, galaxy assembly bias, or both.

\begin{figure}[b!]
    \centering
    \includegraphics[width=1\columnwidth]{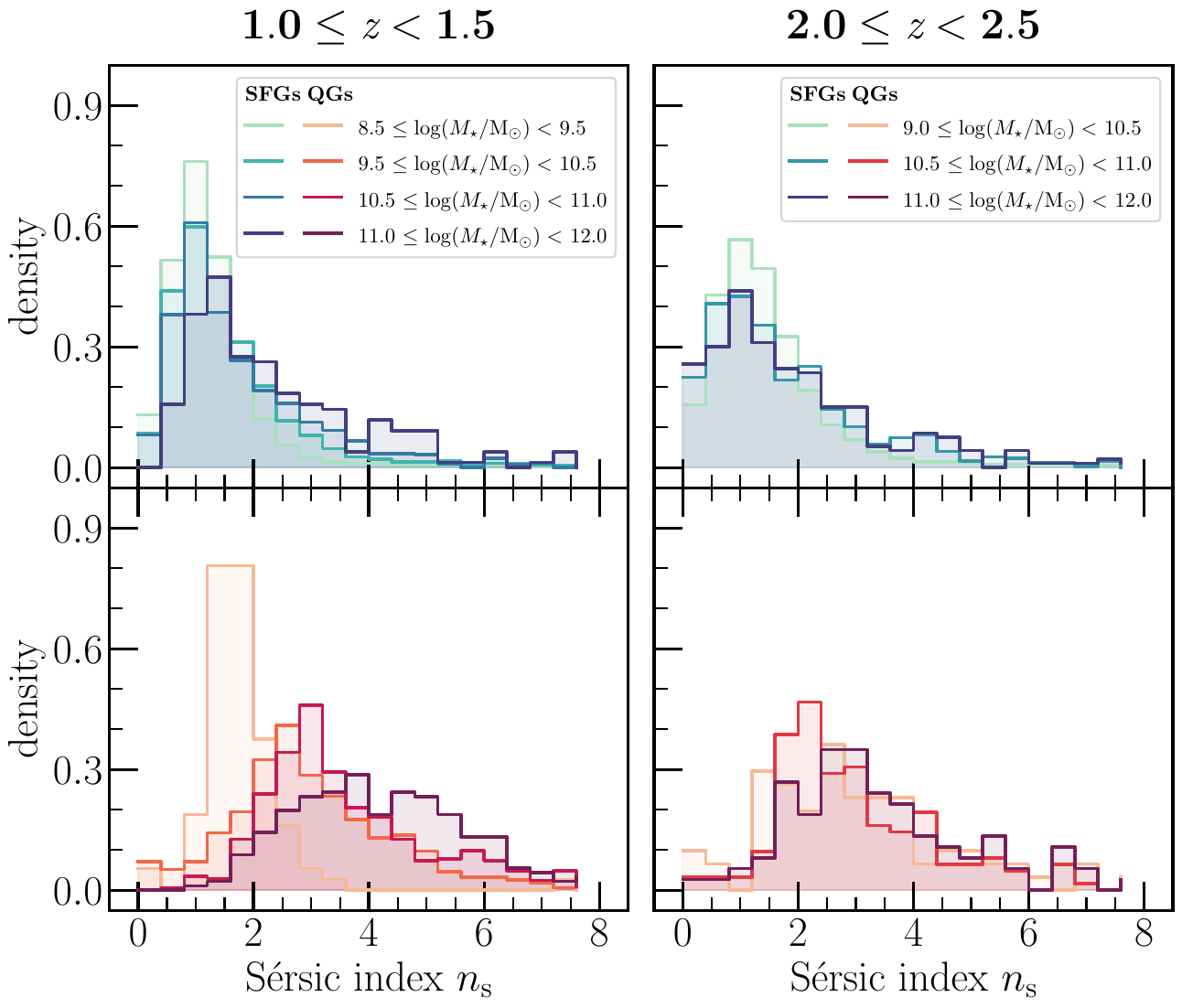}
    \vspace{-5mm}
    \caption{Distribution of the Sérsic index $n_\mathrm{s}$ for SFGs (blue lines) and QGs (red lines), in mass ranges and in two redshift bins.}
    \label{fig:prop_lowmassQT}
\end{figure}

To disentangle the processes involved in their quenching, we show their morphology in \reffig{fig:prop_lowmassQT} via their Sérsic index $n_\mathrm{s}$ (averaged over the $1-5 \,\si{\micro\meter}$ wavelength range). Low-mass QGs exhibit a distinct disky distribution at $z < 2$, which resembles that of SFGs more than that of more massive QGs. This suggests that the quenching mechanism does not significantly disrupt their morphology, thereby ruling out mergers or fly-bys. Instead, it favours ram-pressure stripping, which removes gas from satellites as they enter hot halos \citep[\eg][]{Boselli2006,Boselli2022}, or gas depletion, whereby satellites accreted onto massive hot halos are cut off from fresh cold gas inflows \citep[\eg][]{Hatfield&Jarvis17}.

Therefore, our results provide evidence for the action of environmental quenching at $z \le 2$, whose power increases for lower stellar mass and leaves a signature in galaxies' ACF. While these external quenching mechanisms may depend on halo mass, they might also depend on secondary halo properties or the cosmic web environment at fixed halo mass. Ram-pressure stripping for instance could affect a galaxy as a function of the distance to the halo centre or halo density rather than solely on its mass. 

\subsection{One-halo conformity up to $z \simeq 2$}   \label{subsec:conformity} 

If quenching is influenced by secondary halo properties or the large-scale environment beyond halo mass, the SFRs of galaxies in a common environment (such as a central galaxy with its satellites) may be correlated: this is the conformity phenomenon.

Our results show a one-halo conformity signal: low-mass QGs are more clustered around high-mass QGs than what is measured for star-forming ones, even after controlling for halo mass, and this trend persists up to $z = 2$. A tentative signal is also observed for satellite QGs around central QGs. This is consistent with previous studies, mainly based on SDSS data at low $z$ \citep[\eg][]{Weinmann2006,Kawinwanichakij2016,Hatfield&Jarvis17,Guo2017,Ji18,Zentner2019}; but also in the COSMOS field by \cite{Kaviraj2025}, who finds that half of low-$z$ red dwarfs lie near massive galaxies, arguing this proximity is a stronger quenching driver than distance to filaments, nodes, or local density. Yet, we do not observe this effect beyond the typical halo radius $r_{\rm c} \ge 1\,{\rm Mpc}$, as has been claimed by \cite{Kauffmann2013} or \cite{Calderon2018}. However, we do not claim a complete non-detection of this signal; rather, our survey is not wide enough to accurately probe the two-halo term within the error bars required to identify this low-amplitude signal and overcome the integral constraint.

Thus, central and satellites have correlated histories, either because the mechanism that quenched the central quenches also the neighbouring galaxies (for instance, \citealt{Dashyan2019} show that the AGN feedback from centrals could affect its close neighbours), or because of the environment they share. 

One scenario for the latter is if QGs are hosted by older or more concentrated halos at fixed halo mass \citep[\eg][]{Hearin&Watson2013,Watson2015,Wang2026}, which is a form of assembly bias. For instance, \cite{Oyarzun2024} show that the stellar mass of quiescent centrals correlates with their halo formation time at fixed halo mass; \cite{Wang2022} find that more concentrated halos are more likely to host centrals that formed earlier; and \cite{Zentner2019} provides evidence that models assuming a significant central galaxy assembly bias tied to halo concentration better fit SDSS clustering. In this case, a conformity signal may arise \citep{Hearin2015}: early-formed halos, which tend to be more gas-rich and denser, are more likely to host central galaxies that grew and quenched early, and can more efficiently quench accreted satellites via ram-pressure stripping \citep{Wang&White2012}. This is also consistent with discoveries from \JWST indicating intense star formation which produced very massive galaxies at $z \ge 5$: early-formed halos would then be more likely to host quiescent descendants at lower redshifts. Nonetheless, we find no evidence of any discrepancies related to the age of formation, star formation history or morphology of our quiescent satellites based on the type of their neighbouring central. Further research is necessary to investigate whether assembly bias, if it exists and is responsible for the conformity signal, affects galaxy properties.

The work by \cite{Zu&Mandelbaum2016,Zu&Mandelbaum2018} offers an alternative explanation to conformity requiring no assembly bias: quenching is driven by halo mass alone, but the halo mass function varies with large-scale environment, so that denser regions host more massive halos and thus more QGs. Quenching relies on the ability of the host halo to cool the flowing gas, acting differently on centrals and satellites. This reproduces SDSS conformity signals from galaxy colours, but only at fixed stellar mass $-$ not fixed halo mass $-$ making it unlikely to fully explain our results. One-halo conformity could also be driven by differences in the radial distributions of SFGs and QGs satellites within halos, an effect that requires more information on halo occupation and mass profiles to be properly assessed.

We also note that the signal decreases as $z$ increases, ultimately vanishing at $z \ge 2$. Prior evidence of this evolution has been noted by \cite{Kawinwanichakij2016} up to $z \ge 1.6$; \cite{Hartley2015} reported a 3$\sigma$ one-halo conformity signal in the range of $0.4 \le z < 1.9$; and \cite{Binh2025} show no trace of conformity at $z \sim 3-5$ for massive QGs in COSMOS-Web. Hence, whatever the physical explanation for conformity might be, its impact on galaxy clustering should decline with increasing redshift and disappear after cosmic noon. 

\subsection{Limitations of this work} \label{subsec:limitations}

Detecting conformity is complex because it requires accurately identifying which galaxies are in the same halos and the masses of these halos; otherwise, spurious conformity signals can arise from systematic errors only \citep{Calderon2018}.

\paragraph{Controlling for halo mass.} One of the challenges is to accurately correct for halo mass. Our method relies on the assumption that \UM represents well the spatial distribution of SFGs and QGs across time, along with their properties. This is justified by the fact that \UM reproduces previous clustering measurements by design, and matches the COSMOS-Web SHMR over $z=0.1-14$ \citepalias{Paquereau2025}. Moreover, it is built by correlating star formation histories with halo assembly, which results in the rise of galaxy assembly bias \citep{Behroozi19}: \UM's QGs are associated with older halos that have had no recent accretion, while SFGs are linked to halos that are rapidly accreting, at the same halo mass. Nonetheless, the HMM technique involves removing random objects from the samples, which could dilute or bias the signal. Although beyond the scope of our work, HMM could be tested using weak lensing: mass-matched samples of SFGs and QGs should yield similar lensing measurements if their halo mass distributions are indeed equivalent.

\paragraph{Galaxy properties.} The second source of uncertainty lies in the central/satellite identification: as shown by \cite{Campbell2015} or \cite{Lacerna2018}, misclassifications from group finders can introduce artificial conformity detections. This is why we tested two configurations for the cross-correlations in \refsec{sec:crosscorr}, both in agreement. Furthermore, the purity and completeness of the group catalogue have been examined using galaxy mocks and set to at least 70\%, as discussed by \cite{Toni2025}. The extensive spectroscopic compilation in COSMOS \citep{Khostovan2025} has also facilitated the clear identification of groups, with approximately one-third of groups having spectroscopic confirmation. We tested in more detail the impact of HMM and central/satellite misidentification in \refapp{app:limitations}.

A misclassification of quiescent and star-forming galaxies can also bias the clustering. We confirmed that our findings stay consistent using a $NUVrJ$ criteria instead. Moreover, the 37 photometric bands available in COSMOS-Web, including the deep \JWST imaging, provide rest-frame UV coverage out to $z \sim 4-5$, enabling robust estimates of physical properties.

\paragraph{Statistics.} Finally, we acknowledge the uncertainties inherent to ACF measurements, particularly at $z \ge 2$ where the QGs sample decreases to fewer than 100 objects in each bin, leading to a loss of statistical power. Our analysis also relies on photometric redshift bins, which can bias the clustering signal upward or downward if photo-$z$ are inaccurate, or produce spurious signals due to projection effects (see \citetalias{Paquereau2025} for more details). The relatively small area of COSMOS-Web, compared to wide-field surveys such as SDSS, is also a limitation: \eg \citet{Lim2025} show that the detection amplitude of conformity depends on cosmic variance effects. It also restricts our sensitivity to large-scale conformity signals; for this reason, we focused primarily on the one-halo detection in this study.

\section{Conclusion} \label{sec:conclusion}

We used two-point auto- and cross-correlation functions of quiescent and star-forming galaxies from $z \simeq 5$ to the present-day in the COSMOS-Web survey, to investigate the influence of the environment on galaxy quenching. To isolate effects beyond halo mass, we introduced a procedure matching QGs and SFGs halo mass distributions based on the \textsc{UniverseMachine} model. Our main conclusions are as follows:
\begin{itemize}
    \item QGs are more clustered than SFGs of the same stellar mass across the range $z = 0.1 - 5$, indicating that the environments of the earliest QGs are already distinct from those of their star-forming counterparts. This difference is, however, largely reduced at $z \ge 2$ after removing the effect of halo mass, indicating that QGs most likely live in more massive halos at these early times. Yet, QGs stay more clustered at $z \le 2$ (at small scales), hinting at another effect induced by secondary halo properties or the environment.
    
    \item QGs are more clustered with decreasing stellar mass, where low-mass QGs are the most clustered objects among all populations, up to $z \simeq 2$. SFGs show the opposite trend. These low-mass QGs are likely satellites that underwent rapid environmental quenching of star formation without significantly altering their morphology, as they are mostly discs. Such processes may include ram-pressure stripping or the cessation of new cold gas inflows.
    
    \item Cross-correlations reveal that low-mass (or satellites) QGs are more clustered around high-mass (or centrals) QGs than around star-forming ones or even the overall massive (or central) sample. This is also seen after halo mass matching, with at least a 2$\sigma$ significance on average, but only at low scales ($r \lesssim 1 \,\mathrm{Mpc}$) and up to redshift $z \le 2$. This stands for a detection of one-halo conformity, where galaxies and their neighbouring central have correlated SFRs. This signal may arise either from a quenching mechanism affecting both the central and satellites (\eg AGN feedback), from correlated growth histories prior to infall (\eg via cosmic web processes), from their shared halo environment (\eg in case of a hot and dense halo), or from effects related to secondary halo properties (\eg in case of assembly bias).
\end{itemize}
These results point to the fact that modelling quenching and galaxy correlation functions requires more than halo mass alone. Environmental quenching for low-mass galaxies should be accounted for at $z \le 2$, as well as the proximity to a massive central galaxy (depending on its SFR).

However, limitations in galaxy classification and halo mass estimates call for improved data or statistics to move forward. Halo masses could be better constrained through X-ray emission of groups and clusters \citep[as demonstrated in COSMOS2020 by][]{Toni2024}, or gravitational lensing \citep{Scognamiglio2026}. A natural next step would also be to use halo occupation models \citep[\eg, from][]{Hearin2016b,Hearin2017,Pahwa2017} to investigate conformity scenarios and halo properties in more detail.

\begin{acknowledgements}
Part of this work was presented in a preliminary version in LP's thesis manuscript. 
LP acknowledges the thesis funding from the Centre National d’Etudes Spatiales (CNES) and the Ecole Doctorale Astronomie et Astrophysique d'Ile-de-France. 
The authors acknowledge the funding of the French Agence Nationale de la Recherche for the project iMAGE (grant ANR-22-CE31-0007). 
This work was made possible by utilising the CANDIDE cluster at the Institut d’Astrophysique de Paris, which was funded through grants from the PNCG, CNES, DIM-ACAV, the Euclid Consortium, and the Cosmic Dawn Center (DNRF140) and is maintained by Stephane Rouberol. 
The authors acknowledge the contributions of the COSMOS collaboration. The French contingent of the COSMOS team is partly supported by the CNES. 
\end{acknowledgements}

\bibliographystyle{aa_url}
\bibliography{references}

\begin{thebibliography}{115}
\expandafter\ifx\csname natexlab\endcsname\relax\def\natexlab#1{#1}\fi

\bibitem[{{Alberts} {et~al.}(2024){Alberts}, {Williams}, {Helton}, {Suess},
  {Ji}, {Shivaei}, {Lyu}, {Rieke}, {Baker}, {Bonaventura}, {Bunker},
  {Carniani}, {Charlot}, {Curtis-Lake}, {D'Eugenio}, {Eisenstein}, {de Graaff},
  {Hainline}, {Hausen}, {Johnson}, {Maiolino}, {Parlanti}, {Rieke},
  {Robertson}, {Sun}, {Tacchella}, {Willmer}, \& {Willott}}]{Alberts2024}
{Alberts}, S., {Williams}, C.~C., {Helton}, J.~M., {et~al.} 2024,
  \href{http://dx.doi.org/10.3847/1538-4357/ad66cc}{\color{blue}\apj},
  \href{https://ui.adsabs.harvard.edu/abs/2024ApJ...975...85A}{975, 85}

\bibitem[{{Arango-Toro} {et~al.}(2025){Arango-Toro}, {Ilbert}, {Ciesla},
  {Shuntov}, {Aufort}, {Mercier}, {Laigle}, {Franco}, {Bethermin}, {Le Borgne},
  {Dubois}, {McCracken}, {Paquereau}, {Huertas-Company}, {Kartaltepe}, {Casey},
  {Akins}, {Allen}, {Andika}, {Brinch}, {Drakos}, {Faisst}, {Gozaliasl},
  {Harish}, {Kaminsky}, {Koekemoer}, {Kokorev}, {Liu}, {Magdis}, {Martin},
  {Moutard}, {Rhodes}, {Rich}, {Robertson}, {Sanders}, {Sheth}, {Talia},
  {Toft}, {Tresse}, {Valentino}, {Vijayan}, \& {Weaver}}]{ArangoToro2024}
{Arango-Toro}, R.~C., {Ilbert}, O., {Ciesla}, L., {et~al.} 2025,
  \href{http://dx.doi.org/10.1051/0004-6361/202452519}{\color{blue}\aap},
  \href{https://ui.adsabs.harvard.edu/abs/2025A&A...696A.159A}{696, A159}

\bibitem[{{Arnouts} {et~al.}(2002){Arnouts}, {Moscardini}, {Vanzella},
  {Colombi}, {Cristiani}, {Fontana}, {Giallongo}, {Matarrese}, \&
  {Saracco}}]{Arnouts02_LP}
{Arnouts}, S., {Moscardini}, L., {Vanzella}, E., {et~al.} 2002,
  \href{http://dx.doi.org/10.1046/j.1365-8711.2002.04988.x}{\color{blue}\mnras},
  \href{https://ui.adsabs.harvard.edu/abs/2002MNRAS.329..355A}{329, 355}

\bibitem[{{Arnouts} {et~al.}(2007){Arnouts}, {Walcher}, {Le F{\`e}vre},
  {Zamorani}, {Ilbert}, {Le Brun}, {Pozzetti}, {Bardelli}, {Tresse}, {Zucca},
  {Charlot}, {Lamareille}, {McCracken}, {Bolzonella}, {Iovino}, {Lonsdale},
  {Polletta}, {Surace}, {Bottini}, {Garilli}, {Maccagni}, {Picat},
  {Scaramella}, {Scodeggio}, {Vettolani}, {Zanichelli}, {Adami}, {Cappi},
  {Ciliegi}, {Contini}, {de la Torre}, {Foucaud}, {Franzetti}, {Gavignaud},
  {Guzzo}, {Marano}, {Marinoni}, {Mazure}, {Meneux}, {Merighi}, {Paltani},
  {Pell{\`o}}, {Pollo}, {Radovich}, {Temporin}, \& {Vergani}}]{Arnouts2007}
{Arnouts}, S., {Walcher}, C.~J., {Le F{\`e}vre}, O., {et~al.} 2007,
  \href{http://dx.doi.org/10.1051/0004-6361:20077632}{\color{blue}\aap},
  \href{https://ui.adsabs.harvard.edu/abs/2007A&A...476..137A}{476, 137}

\bibitem[{{Behroozi} {et~al.}(2019){Behroozi}, {Wechsler}, {Hearin}, \&
  {Conroy}}]{Behroozi19}
{Behroozi}, P., {Wechsler}, R.~H., {Hearin}, A.~P., \& {Conroy}, C. 2019,
  \href{http://dx.doi.org/10.1093/mnras/stz1182}{\color{blue}\mnras},
  \href{https://ui.adsabs.harvard.edu/abs/2019MNRAS.488.3143B}{488, 3143}

\bibitem[{{Bellagamba} {et~al.}(2018){Bellagamba}, {Roncarelli}, {Maturi}, \&
  {Moscardini}}]{Bellagamba2018}
{Bellagamba}, F., {Roncarelli}, M., {Maturi}, M., \& {Moscardini}, L. 2018,
  \href{http://dx.doi.org/10.1093/mnras/stx2701}{\color{blue}\mnras},
  \href{https://ui.adsabs.harvard.edu/abs/2018MNRAS.473.5221B}{473, 5221}

\bibitem[{{Berti} {et~al.}(2017){Berti}, {Coil}, {Behroozi}, {Eisenstein},
  {Bray}, {Cool}, \& {Moustakas}}]{Berti2017}
{Berti}, A.~M., {Coil}, A.~L., {Behroozi}, P.~S., {et~al.} 2017,
  \href{http://dx.doi.org/10.3847/1538-4357/834/1/87}{\color{blue}\apj},
  \href{https://ui.adsabs.harvard.edu/abs/2017ApJ...834...87B}{834, 87}

\bibitem[{{Berti} {et~al.}(2021){Berti}, {Coil}, {Hearin}, \&
  {Behroozi}}]{Berti2021}
{Berti}, A.~M., {Coil}, A.~L., {Hearin}, A.~P., \& {Behroozi}, P.~S. 2021,
  \href{http://dx.doi.org/10.3847/1538-3881/abcc6a}{\color{blue}\aj},
  \href{https://ui.adsabs.harvard.edu/abs/2021AJ....161...49B}{161, 49}

\bibitem[{{Berti} {et~al.}(2019){Berti}, {Coil}, {Hearin}, \&
  {Moustakas}}]{Berti2019}
{Berti}, A.~M., {Coil}, A.~L., {Hearin}, A.~P., \& {Moustakas}, J. 2019,
  \href{http://dx.doi.org/10.3847/1538-4357/ab3b5d}{\color{blue}\apj},
  \href{https://ui.adsabs.harvard.edu/abs/2019ApJ...884...76B}{884, 76}

\bibitem[{{Binh} {et~al.}(2025){Binh}, {Long}, {Antwi-Danso}, {Andrews},
  {Toni}, {Champagne}, {Akins}, {Anderson}, {Arango-Toro}, {Casey}, {Cheng},
  {Cooper}, {Drakos}, {Faisst}, {Franco}, {Gammon}, {Hirschmann}, {Ilbert},
  {Kartaltepe}, {Koekemoer}, {Liu}, {Magdis}, {Maturi}, {McCracken},
  {Moscardini}, {Paquereau}, {Rhodes}, {Rich}, {Robertson}, {Shamyati},
  {Shuntov}, \& {Xu}}]{Binh2025}
{Binh}, N., {Long}, A.~S., {Antwi-Danso}, J., {et~al.} 2025,
  \href{https://ui.adsabs.harvard.edu/abs/2025arXiv251214881B}{\href{http://dx.doi.org/10.48550/arXiv.2512.14881}{\color{blue}arXiv
  e-prints}, arXiv:2512.14881}

\bibitem[{{Bluck} {et~al.}(2022){Bluck}, {Maiolino}, {Brownson}, {Conselice},
  {Ellison}, {Piotrowska}, \& {Thorp}}]{Bluck2022}
{Bluck}, A. F.~L., {Maiolino}, R., {Brownson}, S., {et~al.} 2022,
  \href{http://dx.doi.org/10.1051/0004-6361/202142643}{\color{blue}\aap},
  \href{https://ui.adsabs.harvard.edu/abs/2022A&A...659A.160B}{659, A160}

\bibitem[{{Boquien} {et~al.}(2019){Boquien}, {Burgarella}, {Roehlly}, {Buat},
  {Ciesla}, {Corre}, {Inoue}, \& {Salas}}]{Boquien19_CIGALE}
{Boquien}, M., {Burgarella}, D., {Roehlly}, Y., {et~al.} 2019,
  \href{http://dx.doi.org/10.1051/0004-6361/201834156}{\color{blue}\aap},
  \href{https://ui.adsabs.harvard.edu/abs/2019A&A...622A.103B}{622, A103}

\bibitem[{{Boselli} {et~al.}(2022){Boselli}, {Fossati}, \& {Sun}}]{Boselli2022}
{Boselli}, A., {Fossati}, M., \& {Sun}, M. 2022,
  \href{http://dx.doi.org/10.1007/s00159-022-00140-3}{\color{blue}\aapr},
  \href{https://ui.adsabs.harvard.edu/abs/2022A&ARv..30....3B}{30, 3}

\bibitem[{{Boselli} \& {Gavazzi}(2006)}]{Boselli2006}
{Boselli}, A. \& {Gavazzi}, G. 2006,
  \href{http://dx.doi.org/10.1086/500691}{\color{blue}\pasp},
  \href{https://ui.adsabs.harvard.edu/abs/2006PASP..118..517B}{118, 517}

\bibitem[{{Bower} {et~al.}(2006){Bower}, {Benson}, {Malbon}, {Helly}, {Frenk},
  {Baugh}, {Cole}, \& {Lacey}}]{Bower2006}
{Bower}, R.~G., {Benson}, A.~J., {Malbon}, R., {et~al.} 2006,
  \href{http://dx.doi.org/10.1111/j.1365-2966.2006.10519.x}{\color{blue}\mnras},
  \href{https://ui.adsabs.harvard.edu/abs/2006MNRAS.370..645B}{370, 645}

\bibitem[{{Calderon} {et~al.}(2018){Calderon}, {Berlind}, \&
  {Sinha}}]{Calderon2018}
{Calderon}, V.~F., {Berlind}, A.~A., \& {Sinha}, M. 2018,
  \href{http://dx.doi.org/10.1093/mnras/sty2000}{\color{blue}\mnras},
  \href{https://ui.adsabs.harvard.edu/abs/2018MNRAS.480.2031C}{480, 2031}

\bibitem[{{Campbell} {et~al.}(2015){Campbell}, {van den Bosch}, {Hearin},
  {Padmanabhan}, {Berlind}, {Mo}, {Tinker}, \& {Yang}}]{Campbell2015}
{Campbell}, D., {van den Bosch}, F.~C., {Hearin}, A., {et~al.} 2015,
  \href{http://dx.doi.org/10.1093/mnras/stv1091}{\color{blue}\mnras},
  \href{https://ui.adsabs.harvard.edu/abs/2015MNRAS.452..444C}{452, 444}

\bibitem[{{Carnall} {et~al.}(2023){Carnall}, {McLeod}, {McLure}, {Dunlop},
  {Begley}, {Cullen}, {Donnan}, {Hamadouche}, {Jewell}, {Jones}, {Pollock}, \&
  {Wild}}]{Carnall2023}
{Carnall}, A.~C., {McLeod}, D.~J., {McLure}, R.~J., {et~al.} 2023,
  \href{http://dx.doi.org/10.1093/mnras/stad369}{\color{blue}\mnras},
  \href{https://ui.adsabs.harvard.edu/abs/2023MNRAS.520.3974C}{520, 3974}

\bibitem[{{Carnall} {et~al.}(2018){Carnall}, {McLure}, {Dunlop}, \&
  {Dav{\'e}}}]{Carnall2018}
{Carnall}, A.~C., {McLure}, R.~J., {Dunlop}, J.~S., \& {Dav{\'e}}, R. 2018,
  \href{http://dx.doi.org/10.1093/mnras/sty2169}{\color{blue}\mnras},
  \href{https://ui.adsabs.harvard.edu/abs/2018MNRAS.480.4379C}{480, 4379}

\bibitem[{{Casey} {et~al.}(2023){Casey}, {Kartaltepe}, {Drakos}, {Franco},
  {Harish}, {Paquereau}, {Ilbert}, {Rose}, {Cox}, {Nightingale}, {Robertson},
  {Silverman}, {Koekemoer}, {Massey}, {McCracken}, {Rhodes}, {Akins}, {Allen},
  {Amvrosiadis}, {Arango-Toro}, {Bagley}, {Bongiorno}, {Capak}, {Champagne},
  {Chartab}, {Ch{\'a}vez Ortiz}, {Chworowsky}, {Cooke}, {Cooper}, {Darvish},
  {Ding}, {Faisst}, {Finkelstein}, {Fujimoto}, {Gentile}, {Gillman}, {Gould},
  {Gozaliasl}, {Hayward}, {He}, {Hemmati}, {Hirschmann}, {Jahnke}, {Jin},
  {Khostovan}, {Kokorev}, {Lambrides}, {Laigle}, {Larson}, {Leung}, {Liu},
  {Liaudat}, {Long}, {Magdis}, {Mahler}, {Mainieri}, {Manning}, {Maraston},
  {Martin}, {McCleary}, {McKinney}, {McPartland}, {Mobasher}, {Pattnaik},
  {Renzini}, {Rich}, {Sanders}, {Sattari}, {Scognamiglio}, {Scoville}, {Sheth},
  {Shuntov}, {Sparre}, {Suzuki}, {Talia}, {Toft}, {Trakhtenbrot}, {Urry},
  {Valentino}, {Vanderhoof}, {Vardoulaki}, {Weaver}, {Whitaker}, {Wilkins},
  {Yang}, \& {Zavala}}]{Casey23_CWeb}
{Casey}, C.~M., {Kartaltepe}, J.~S., {Drakos}, N.~E., {et~al.} 2023,
  \href{http://dx.doi.org/10.3847/1538-4357/acc2bc}{\color{blue}\apj},
  \href{https://ui.adsabs.harvard.edu/abs/2023ApJ...954...31C}{954, 31}

\bibitem[{{Cattaneo} {et~al.}(2006){Cattaneo}, {Dekel}, {Devriendt},
  {Guiderdoni}, \& {Blaizot}}]{Cattaneo2006}
{Cattaneo}, A., {Dekel}, A., {Devriendt}, J., {Guiderdoni}, B., \& {Blaizot},
  J. 2006,
  \href{http://dx.doi.org/10.1111/j.1365-2966.2006.10608.x}{\color{blue}\mnras},
  \href{https://ui.adsabs.harvard.edu/abs/2006MNRAS.370.1651C}{370, 1651}

\bibitem[{{Chabrier}(2003)}]{Chabrier03}
{Chabrier}, G. 2003,
  \href{http://dx.doi.org/10.1086/376392}{\color{blue}\pasp},
  \href{https://ui.adsabs.harvard.edu/abs/2003PASP..115..763C}{115, 763}

\bibitem[{{Coil} {et~al.}(2017){Coil}, {Mendez}, {Eisenstein}, \&
  {Moustakas}}]{Coil2017}
{Coil}, A.~L., {Mendez}, A.~J., {Eisenstein}, D.~J., \& {Moustakas}, J. 2017,
  \href{http://dx.doi.org/10.3847/1538-4357/aa63ec}{\color{blue}\apj},
  \href{https://ui.adsabs.harvard.edu/abs/2017ApJ...838...87C}{838, 87}

\bibitem[{{Croton} {et~al.}(2007){Croton}, {Gao}, \& {White}}]{Croton2007}
{Croton}, D.~J., {Gao}, L., \& {White}, S. D.~M. 2007,
  \href{http://dx.doi.org/10.1111/j.1365-2966.2006.11230.x}{\color{blue}\mnras},
  \href{https://ui.adsabs.harvard.edu/abs/2007MNRAS.374.1303C}{374, 1303}

\bibitem[{{Croton} {et~al.}(2006){Croton}, {Springel}, {White}, {De Lucia},
  {Frenk}, {Gao}, {Jenkins}, {Kauffmann}, {Navarro}, \& {Yoshida}}]{Croton2006}
{Croton}, D.~J., {Springel}, V., {White}, S. D.~M., {et~al.} 2006,
  \href{http://dx.doi.org/10.1111/j.1365-2966.2005.09675.x}{\color{blue}\mnras},
  \href{https://ui.adsabs.harvard.edu/abs/2006MNRAS.365...11C}{365, 11}

\bibitem[{{Darvish} {et~al.}(2014){Darvish}, {Sobral}, {Mobasher}, {Scoville},
  {Best}, {Sales}, \& {Smail}}]{Darvish2014}
{Darvish}, B., {Sobral}, D., {Mobasher}, B., {et~al.} 2014,
  \href{http://dx.doi.org/10.1088/0004-637X/796/1/51}{\color{blue}\apj},
  \href{https://ui.adsabs.harvard.edu/abs/2014ApJ...796...51D}{796, 51}

\bibitem[{{Dashyan} {et~al.}(2019){Dashyan}, {Choi}, {Somerville}, {Naab},
  {Quirk}, {Hirschmann}, \& {Ostriker}}]{Dashyan2019}
{Dashyan}, G., {Choi}, E., {Somerville}, R.~S., {et~al.} 2019,
  \href{http://dx.doi.org/10.1093/mnras/stz1697}{\color{blue}\mnras},
  \href{https://ui.adsabs.harvard.edu/abs/2019MNRAS.487.5889D}{487, 5889}

\bibitem[{{de Graaff} {et~al.}(2025){de Graaff}, {Setton}, {Brammer}, {Cutler},
  {Suess}, {Labb{\'e}}, {Leja}, {Weibel}, {Maseda}, {Whitaker}, {Bezanson},
  {Boogaard}, {Cleri}, {De Lucia}, {Franx}, {Greene}, {Hirschmann}, {Matthee},
  {McConachie}, {Naidu}, {Oesch}, {Price}, {Rix}, {Valentino}, {Wang}, \&
  {Williams}}]{deGraaff2025}
{de Graaff}, A., {Setton}, D.~J., {Brammer}, G., {et~al.} 2025,
  \href{http://dx.doi.org/10.1038/s41550-024-02424-3}{\color{blue}Nature
  Astronomy}, \href{https://ui.adsabs.harvard.edu/abs/2025NatAs...9..280D}{9,
  280}

\bibitem[{{Dekel} \& {Birnboim}(2006)}]{Dekel2006}
{Dekel}, A. \& {Birnboim}, Y. 2006,
  \href{http://dx.doi.org/10.1111/j.1365-2966.2006.10145.x}{\color{blue}\mnras},
  \href{https://ui.adsabs.harvard.edu/abs/2006MNRAS.368....2D}{368, 2}

\bibitem[{{Dressler}(1980)}]{Dressler1980}
{Dressler}, A. 1980, \href{http://dx.doi.org/10.1086/157753}{\color{blue}\apj},
  \href{https://ui.adsabs.harvard.edu/abs/1980ApJ...236..351D}{236, 351}

\bibitem[{{Fang} {et~al.}(2018){Fang}, {Faber}, {Koo}, {Rodr{\'\i}guez-Puebla},
  {Guo}, {Barro}, {Behroozi}, {Brammer}, {Chen}, {Dekel}, {Ferguson},
  {Gawiser}, {Giavalisco}, {Kartaltepe}, {Kocevski}, {Koekemoer}, {McGrath},
  {McIntosh}, {Newman}, {Pacifici}, {Pandya}, {P{\'e}rez-Gonz{\'a}lez},
  {Primack}, {Salmon}, {Trump}, {Weiner}, {Willner}, {Acquaviva}, {Dahlen},
  {Finkelstein}, {Finlator}, {Fontana}, {Galametz}, {Grogin}, {Gruetzbauch},
  {Johnson}, {Mobasher}, {Papovich}, {Pforr}, {Salvato}, {Santini}, {van der
  Wel}, {Wiklind}, \& {Wuyts}}]{Fang2018}
{Fang}, J.~J., {Faber}, S.~M., {Koo}, D.~C., {et~al.} 2018,
  \href{http://dx.doi.org/10.3847/1538-4357/aabcba}{\color{blue}\apj},
  \href{https://ui.adsabs.harvard.edu/abs/2018ApJ...858..100F}{858, 100}

\bibitem[{{Farouki} \& {Shapiro}(1981)}]{Farouki&Shapiro1981}
{Farouki}, R. \& {Shapiro}, S.~L. 1981,
  \href{http://dx.doi.org/10.1086/158563}{\color{blue}\apj},
  \href{https://ui.adsabs.harvard.edu/abs/1981ApJ...243...32F}{243, 32}

\bibitem[{{Fontana} {et~al.}(2009){Fontana}, {Santini}, {Grazian},
  {Pentericci}, {Fiore}, {Castellano}, {Giallongo}, {Menci}, {Salimbeni},
  {Cristiani}, {Nonino}, \& {Vanzella}}]{Fontana2009}
{Fontana}, A., {Santini}, P., {Grazian}, A., {et~al.} 2009,
  \href{http://dx.doi.org/10.1051/0004-6361/200911650}{\color{blue}\aap},
  \href{https://ui.adsabs.harvard.edu/abs/2009A&A...501...15F}{501, 15}

\bibitem[{{Geha} {et~al.}(2012){Geha}, {Blanton}, {Yan}, \&
  {Tinker}}]{Geha2012}
{Geha}, M., {Blanton}, M.~R., {Yan}, R., \& {Tinker}, J.~L. 2012,
  \href{http://dx.doi.org/10.1088/0004-637X/757/1/85}{\color{blue}\apj},
  \href{https://ui.adsabs.harvard.edu/abs/2012ApJ...757...85G}{757, 85}

\bibitem[{{Groth} \& {Peebles}(1977)}]{Groth&Peebles1977}
{Groth}, E.~J. \& {Peebles}, P.~J.~E. 1977,
  \href{http://dx.doi.org/10.1086/155588}{\color{blue}\apj},
  \href{https://ui.adsabs.harvard.edu/abs/1977ApJ...217..385G}{217, 385}

\bibitem[{{Gunn} \& {Gott}(1972)}]{Gunn&Gott1972}
{Gunn}, J.~E. \& {Gott}, III, J.~R. 1972,
  \href{http://dx.doi.org/10.1086/151605}{\color{blue}\apj},
  \href{https://ui.adsabs.harvard.edu/abs/1972ApJ...176....1G}{176, 1}

\bibitem[{{Guo} {et~al.}(2017){Guo}, {Bell}, {Lu}, {Koo}, {Faber}, {Koekemoer},
  {Kurczynski}, {Lee}, {Papovich}, {Chen}, {Dekel}, {Ferguson}, {Fontana},
  {Giavalisco}, {Kocevski}, {Nayyeri}, {P{\'e}rez-Gonz{\'a}lez}, {Pforr},
  {Rodr{\'\i}guez-Puebla}, \& {Santini}}]{Guo2017}
{Guo}, Y., {Bell}, E.~F., {Lu}, Y., {et~al.} 2017,
  \href{http://dx.doi.org/10.3847/2041-8213/aa70e9}{\color{blue}\apjl},
  \href{https://ui.adsabs.harvard.edu/abs/2017ApJ...841L..22G}{841, L22}

\bibitem[{{Hartley} {et~al.}(2015){Hartley}, {Conselice}, {Mortlock},
  {Foucaud}, \& {Simpson}}]{Hartley2015}
{Hartley}, W.~G., {Conselice}, C.~J., {Mortlock}, A., {Foucaud}, S., \&
  {Simpson}, C. 2015,
  \href{http://dx.doi.org/10.1093/mnras/stv972}{\color{blue}\mnras},
  \href{https://ui.adsabs.harvard.edu/abs/2015MNRAS.451.1613H}{451, 1613}

\bibitem[{{Hatamnia} {et~al.}(2026){Hatamnia}, {Mobasher}, {Taamoli},
  {Kartaltepe}, {Casey}, {Akins}, {Brinch}, {Chartab}, {Drakos}, {Faisst},
  {Finkelstein}, {Franco}, {Giddings}, {Gozaliasl}, {Hadi}, {Haghjoo},
  {Harish}, {Ilbert}, {Jablonka}, {Jin}, {Khostovan}, {Koekemoer}, {Laishram},
  {Liu}, {Maturi}, {McCracken}, {Martin}, {Moscardini}, {Scognamiglio},
  {Shuntov}, {Toni}, {de la Vega}, {Weaver}, \& {Yang}}]{Hatamnia2026}
{Hatamnia}, H., {Mobasher}, B., {Taamoli}, S., {et~al.} 2026,
  \href{http://dx.doi.org/10.3847/1538-4357/ae5bac}{\color{blue}\apj},
  \href{https://ui.adsabs.harvard.edu/abs/2026ApJ..1002..192H}{1002, 192}

\bibitem[{{Hatfield} \& {Jarvis}(2017)}]{Hatfield&Jarvis17}
{Hatfield}, P.~W. \& {Jarvis}, M.~J. 2017,
  \href{http://dx.doi.org/10.1093/mnras/stx2155}{\color{blue}\mnras},
  \href{https://ui.adsabs.harvard.edu/abs/2017MNRAS.472.3570H}{472, 3570}

\bibitem[{{Hearin} {et~al.}(2016{\natexlab{a}}){Hearin}, {Behroozi}, \& {van
  den Bosch}}]{Hearin2016a}
{Hearin}, A.~P., {Behroozi}, P.~S., \& {van den Bosch}, F.~C.
  2016{\natexlab{a}},
  \href{http://dx.doi.org/10.1093/mnras/stw1462}{\color{blue}\mnras},
  \href{https://ui.adsabs.harvard.edu/abs/2016MNRAS.461.2135H}{461, 2135}

\bibitem[{{Hearin} {et~al.}(2017){Hearin}, {Campbell}, {Tollerud}, {Behroozi},
  {Diemer}, {Goldbaum}, {Jennings}, {Leauthaud}, {Mao}, {More}, {Parejko},
  {Sinha}, {Sip{\"o}cz}, \& {Zentner}}]{Hearin2017}
{Hearin}, A.~P., {Campbell}, D., {Tollerud}, E., {et~al.} 2017,
  \href{http://dx.doi.org/10.3847/1538-3881/aa859f}{\color{blue}\aj},
  \href{https://ui.adsabs.harvard.edu/abs/2017AJ....154..190H}{154, 190}

\bibitem[{{Hearin} \& {Watson}(2013)}]{Hearin&Watson2013}
{Hearin}, A.~P. \& {Watson}, D.~F. 2013,
  \href{http://dx.doi.org/10.1093/mnras/stt1374}{\color{blue}\mnras},
  \href{https://ui.adsabs.harvard.edu/abs/2013MNRAS.435.1313H}{435, 1313}

\bibitem[{{Hearin} {et~al.}(2015){Hearin}, {Watson}, \& {van den
  Bosch}}]{Hearin2015}
{Hearin}, A.~P., {Watson}, D.~F., \& {van den Bosch}, F.~C. 2015,
  \href{http://dx.doi.org/10.1093/mnras/stv1358}{\color{blue}\mnras},
  \href{https://ui.adsabs.harvard.edu/abs/2015MNRAS.452.1958H}{452, 1958}

\bibitem[{{Hearin} {et~al.}(2016{\natexlab{b}}){Hearin}, {Zentner}, {van den
  Bosch}, {Campbell}, \& {Tollerud}}]{Hearin2016b}
{Hearin}, A.~P., {Zentner}, A.~R., {van den Bosch}, F.~C., {Campbell}, D., \&
  {Tollerud}, E. 2016{\natexlab{b}},
  \href{http://dx.doi.org/10.1093/mnras/stw840}{\color{blue}\mnras},
  \href{https://ui.adsabs.harvard.edu/abs/2016MNRAS.460.2552H}{460, 2552}

\bibitem[{{Ilbert} {et~al.}(2006){Ilbert}, {Arnouts}, {McCracken},
  {Bolzonella}, {Bertin}, {Le F{\`e}vre}, {Mellier}, {Zamorani}, {Pell{\`o}},
  {Iovino}, {Tresse}, {Le Brun}, {Bottini}, {Garilli}, {Maccagni}, {Picat},
  {Scaramella}, {Scodeggio}, {Vettolani}, {Zanichelli}, {Adami}, {Bardelli},
  {Cappi}, {Charlot}, {Ciliegi}, {Contini}, {Cucciati}, {Foucaud}, {Franzetti},
  {Gavignaud}, {Guzzo}, {Marano}, {Marinoni}, {Mazure}, {Meneux}, {Merighi},
  {Paltani}, {Pollo}, {Pozzetti}, {Radovich}, {Zucca}, {Bondi}, {Bongiorno},
  {Busarello}, {de La Torre}, {Gregorini}, {Lamareille}, {Mathez}, {Merluzzi},
  {Ripepi}, {Rizzo}, \& {Vergani}}]{Ilbert06_LP}
{Ilbert}, O., {Arnouts}, S., {McCracken}, H.~J., {et~al.} 2006,
  \href{http://dx.doi.org/10.1051/0004-6361:20065138}{\color{blue}\aap},
  \href{https://ui.adsabs.harvard.edu/abs/2006A&A...457..841I}{457, 841}

\bibitem[{{Ilbert} {et~al.}(2013){Ilbert}, {McCracken}, {Le F{\`e}vre},
  {Capak}, {Dunlop}, {Karim}, {Renzini}, {Caputi}, {Boissier}, {Arnouts},
  {Aussel}, {Comparat}, {Guo}, {Hudelot}, {Kartaltepe}, {Kneib}, {Krogager},
  {Le Floc'h}, {Lilly}, {Mellier}, {Milvang-Jensen}, {Moutard}, {Onodera},
  {Richard}, {Salvato}, {Sanders}, {Scoville}, {Silverman}, {Taniguchi},
  {Tasca}, {Thomas}, {Toft}, {Tresse}, {Vergani}, {Wolk}, \&
  {Zirm}}]{Ilbert2013}
{Ilbert}, O., {McCracken}, H.~J., {Le F{\`e}vre}, O., {et~al.} 2013,
  \href{http://dx.doi.org/10.1051/0004-6361/201321100}{\color{blue}\aap},
  \href{https://ui.adsabs.harvard.edu/abs/2013A&A...556A..55I}{556, A55}

\bibitem[{{Jarvis} {et~al.}(2004){Jarvis}, {Bernstein}, \&
  {Jain}}]{Jarvis_treecorr}
{Jarvis}, M., {Bernstein}, G., \& {Jain}, B. 2004,
  \href{http://dx.doi.org/10.1111/j.1365-2966.2004.07926.x}{\color{blue}\mnras},
  \href{https://ui.adsabs.harvard.edu/abs/2004MNRAS.352..338J}{352, 338}

\bibitem[{{Jego} {et~al.}(2026){Jego}, {B{\'e}thermin}, {Kraljic}, {Laigle},
  {Wang}, {La Marca}, {Ilbert}, {Akins}, {Casey}, {Leroy}, {Hadi},
  {Kartaltepe}, {Koekemoer}, {McCracken}, {Paquereau}, {Rhodes}, {Robertson},
  {Shuntov}, {Toni}, \& {Xu}}]{Jego2026}
{Jego}, B., {B{\'e}thermin}, M., {Kraljic}, K., {et~al.} 2026,
  \href{https://ui.adsabs.harvard.edu/abs/2026arXiv260221890J}{\href{http://dx.doi.org/10.48550/arXiv.2602.21890}{\color{blue}arXiv
  e-prints}, arXiv:2602.21890}

\bibitem[{{Ji} {et~al.}(2018){Ji}, {Giavalisco}, {Williams}, {Faber},
  {Ferguson}, {Guo}, {Liu}, \& {Lee}}]{Ji18}
{Ji}, Z., {Giavalisco}, M., {Williams}, C.~C., {et~al.} 2018,
  \href{http://dx.doi.org/10.3847/1538-4357/aacc2c}{\color{blue}\apj},
  \href{https://ui.adsabs.harvard.edu/abs/2018ApJ...862..135J}{862, 135}

\bibitem[{{Kauffmann} {et~al.}(2013){Kauffmann}, {Li}, {Zhang}, \&
  {Weinmann}}]{Kauffmann2013}
{Kauffmann}, G., {Li}, C., {Zhang}, W., \& {Weinmann}, S. 2013,
  \href{http://dx.doi.org/10.1093/mnras/stt007}{\color{blue}\mnras},
  \href{https://ui.adsabs.harvard.edu/abs/2013MNRAS.430.1447K}{430, 1447}

\bibitem[{{Kaviraj} {et~al.}(2025){Kaviraj}, {Lazar}, {Watkins}, {Laigle},
  {Martin}, \& {Jackson}}]{Kaviraj2025}
{Kaviraj}, S., {Lazar}, I., {Watkins}, A.~E., {et~al.} 2025,
  \href{http://dx.doi.org/10.1093/mnras/staf233}{\color{blue}\mnras},
  \href{https://ui.adsabs.harvard.edu/abs/2025MNRAS.538..153K}{538, 153}

\bibitem[{{Kawinwanichakij} {et~al.}(2016){Kawinwanichakij}, {Quadri},
  {Papovich}, {Kacprzak}, {Labb{\'e}}, {Spitler}, {Straatman}, {Tran}, {Allen},
  {Behroozi}, {Cowley}, {Dekel}, {Glazebrook}, {Hartley}, {Kelson}, {Koo},
  {Lee}, {Lu}, {Nanayakkara}, {Persson}, {Primack}, {Tilvi}, {Tomczak}, \& {van
  Dokkum}}]{Kawinwanichakij2016}
{Kawinwanichakij}, L., {Quadri}, R.~F., {Papovich}, C., {et~al.} 2016,
  \href{http://dx.doi.org/10.3847/0004-637X/817/1/9}{\color{blue}\apj},
  \href{https://ui.adsabs.harvard.edu/abs/2016ApJ...817....9K}{817, 9}

\bibitem[{{Khostovan} {et~al.}(2026){Khostovan}, {Kartaltepe}, {Salvato},
  {Ilbert}, {Casey}, {Algera}, {Antwi-Danso}, {Battisti}, {Brinch}, {Brusa},
  {Calabr{\`o}}, {Capak}, {Chartab}, {Cooper}, {Cox}, {Darvish}, {Drakos},
  {Faisst}, {George}, {Gozaliasl}, {Harish}, {Hasinger}, {Hatamnia}, {Iovino},
  {Jin}, {Kashino}, {Koekemoer}, {Laishram}, {Lee}, {Lertprasertpong}, {Lilly},
  {Liu}, {Masters}, {Mobasher}, {Nagao}, {Onodera}, {Peng}, {Sanders},
  {Sanders}, {Sattari}, {Scoville}, {Shah}, {Silverman}, {Suzuki}, {Taamoli},
  {Tanaka}, {Tasca}, {Toft}, {Toni}, {Trakhtenbrot}, {Trump}, {Vaccari},
  {Valentino}, {Vanderhoof}, {Weaver}, {Yun}, \& {Zavala}}]{Khostovan2025}
{Khostovan}, A.~A., {Kartaltepe}, J.~S., {Salvato}, M., {et~al.} 2026,
  \href{http://dx.doi.org/10.3847/1538-4365/ae1cb9}{\color{blue}\apjs},
  \href{https://ui.adsabs.harvard.edu/abs/2026ApJS..282....6K}{282, 6}

\bibitem[{{Knobel} {et~al.}(2015){Knobel}, {Lilly}, {Woo}, \&
  {Kova{\v{c}}}}]{Knobel2015}
{Knobel}, C., {Lilly}, S.~J., {Woo}, J., \& {Kova{\v{c}}}, K. 2015,
  \href{http://dx.doi.org/10.1088/0004-637X/800/1/24}{\color{blue}\apj},
  \href{https://ui.adsabs.harvard.edu/abs/2015ApJ...800...24K}{800, 24}

\bibitem[{{Labb{\'e}} {et~al.}(2005){Labb{\'e}}, {Huang}, {Franx}, {Rudnick},
  {Barmby}, {Daddi}, {van Dokkum}, {Fazio}, {F{\"o}rster Schreiber},
  {Moorwood}, {Rix}, {R{\"o}ttgering}, {Trujillo}, \& {van der
  Werf}}]{Labbe2005}
{Labb{\'e}}, I., {Huang}, J., {Franx}, M., {et~al.} 2005,
  \href{http://dx.doi.org/10.1086/430700}{\color{blue}\apjl},
  \href{https://ui.adsabs.harvard.edu/abs/2005ApJ...624L..81L}{624, L81}

\bibitem[{{Lacerna} {et~al.}(2018){Lacerna}, {Contreras}, {Gonz{\'a}lez},
  {Padilla}, \& {Gonzalez-Perez}}]{Lacerna2018}
{Lacerna}, I., {Contreras}, S., {Gonz{\'a}lez}, R.~E., {Padilla}, N., \&
  {Gonzalez-Perez}, V. 2018,
  \href{http://dx.doi.org/10.1093/mnras/stx3253}{\color{blue}\mnras},
  \href{https://ui.adsabs.harvard.edu/abs/2018MNRAS.475.1177L}{475, 1177}

\bibitem[{{Laigle} {et~al.}(2018){Laigle}, {Pichon}, {Arnouts}, {McCracken},
  {Dubois}, {Devriendt}, {Slyz}, {Le Borgne}, {Benoit-L{\'e}vy}, {Hwang},
  {Ilbert}, {Kraljic}, {Malavasi}, {Park}, \& {Vibert}}]{Laigle2018}
{Laigle}, C., {Pichon}, C., {Arnouts}, S., {et~al.} 2018,
  \href{http://dx.doi.org/10.1093/mnras/stx3055}{\color{blue}\mnras},
  \href{https://ui.adsabs.harvard.edu/abs/2018MNRAS.474.5437L}{474, 5437}

\bibitem[{{Landy} \& {Szalay}(1993)}]{LandySzalay93}
{Landy}, S.~D. \& {Szalay}, A.~S. 1993,
  \href{http://dx.doi.org/10.1086/172900}{\color{blue}\apj},
  \href{https://ui.adsabs.harvard.edu/abs/1993ApJ...412...64L}{412, 64}

\bibitem[{{Li} {et~al.}(2025){Li}, {Conselice}, {Sarron}, {Harvey}, {Austin},
  {Adams}, {Trussler}, {Duan}, {Ferreira}, {Westcott}, {Harris}, {Dole},
  {Grogin}, {Frye}, {Koekemoer}, {Robertson}, {Windhorst}, {Polletta}, {Hathi},
  \& {Jansen}}]{Li2025}
{Li}, Q., {Conselice}, C.~J., {Sarron}, F., {et~al.} 2025,
  \href{http://dx.doi.org/10.1093/mnras/staf543}{\color{blue}\mnras},
  \href{https://ui.adsabs.harvard.edu/abs/2025MNRAS.539.1796L}{539, 1796}

\bibitem[{{Lim} {et~al.}(2025){Lim}, {Tacchella}, {Maiolino}, {Lovell}, \&
  {Schaye}}]{Lim2025}
{Lim}, S., {Tacchella}, S., {Maiolino}, R., {Lovell}, C.~C., \& {Schaye}, J.
  2025,
  \href{https://ui.adsabs.harvard.edu/abs/2025arXiv251109618L}{\href{http://dx.doi.org/10.48550/arXiv.2511.09618}{\color{blue}arXiv
  e-prints}, arXiv:2511.09618}

\bibitem[{{Long} {et~al.}(2024){Long}, {Antwi-Danso}, {Lambrides}, {Lovell},
  {de la Vega}, {Valentino}, {Zavala}, {Casey}, {Wilkins}, {Yung}, {Arrabal
  Haro}, {Bagley}, {Bisigello}, {Chworowsky}, {Cooper}, {Cooper}, {Cooray},
  {Croton}, {Dickinson}, {Finkelstein}, {Franco}, {Gould}, {Hirschmann},
  {Hutchison}, {Kartaltepe}, {Kocevski}, {Koekemoer}, {Lucas}, {McKinney},
  {Nere}, {Papovich}, {P{\'e}rez-Gonz{\'a}lez}, {Pirzkal}, \&
  {Santini}}]{Long2024}
{Long}, A.~S., {Antwi-Danso}, J., {Lambrides}, E.~L., {et~al.} 2024,
  \href{http://dx.doi.org/10.3847/1538-4357/ad4cea}{\color{blue}\apj},
  \href{https://ui.adsabs.harvard.edu/abs/2024ApJ...970...68L}{970, 68}

\bibitem[{{Lyu} {et~al.}(2023){Lyu}, {Peng}, {Jing}, {Yang}, {Ho}, {Renzini},
  {Wang}, {Wang}, {Xu}, {Zhao}, {Dou}, {Gu}, {Maiolino}, {Mannucci}, \&
  {Yuan}}]{Lyu2023}
{Lyu}, C., {Peng}, Y., {Jing}, Y., {et~al.} 2023,
  \href{http://dx.doi.org/10.3847/1538-4357/ad036b}{\color{blue}\apj},
  \href{https://ui.adsabs.harvard.edu/abs/2023ApJ...959....5L}{959, 5}

\bibitem[{{McCracken} {et~al.}(2008){McCracken}, {Ilbert}, {Mellier}, {Bertin},
  {Guzzo}, {Arnouts}, {Le F{\`e}vre}, \& {Zamorani}}]{McCracken2008}
{McCracken}, H.~J., {Ilbert}, O., {Mellier}, Y., {et~al.} 2008,
  \href{http://dx.doi.org/10.1051/0004-6361:20078636}{\color{blue}\aap},
  \href{https://ui.adsabs.harvard.edu/abs/2008A&A...479..321M}{479, 321}

\bibitem[{{More} {et~al.}(2011){More}, {van den Bosch}, {Cacciato}, {Skibba},
  {Mo}, \& {Yang}}]{More2011}
{More}, S., {van den Bosch}, F.~C., {Cacciato}, M., {et~al.} 2011,
  \href{http://dx.doi.org/10.1111/j.1365-2966.2010.17436.x}{\color{blue}\mnras},
  \href{https://ui.adsabs.harvard.edu/abs/2011MNRAS.410..210M}{410, 210}

\bibitem[{{Moutard} {et~al.}(2018){Moutard}, {Sawicki}, {Arnouts}, {Golob},
  {Malavasi}, {Adami}, {Coupon}, \& {Ilbert}}]{Moutard2018}
{Moutard}, T., {Sawicki}, M., {Arnouts}, S., {et~al.} 2018,
  \href{http://dx.doi.org/10.1093/mnras/sty1543}{\color{blue}\mnras},
  \href{https://ui.adsabs.harvard.edu/abs/2018MNRAS.479.2147M}{479, 2147}

\bibitem[{{Navarro} {et~al.}(1997){Navarro}, {Frenk}, \& {White}}]{NFW1997}
{Navarro}, J.~F., {Frenk}, C.~S., \& {White}, S. D.~M. 1997,
  \href{http://dx.doi.org/10.1086/304888}{\color{blue}\apj},
  \href{https://ui.adsabs.harvard.edu/abs/1997ApJ...490..493N}{490, 493}

\bibitem[{{Oke} \& {Gunn}(1983)}]{OkeGunn1983_ABmag}
{Oke}, J.~B. \& {Gunn}, J.~E. 1983,
  \href{http://dx.doi.org/10.1086/160817}{\color{blue}\apj},
  \href{https://ui.adsabs.harvard.edu/abs/1983ApJ...266..713O}{266, 713}

\bibitem[{{Oyarz{\'u}n} {et~al.}(2024){Oyarz{\'u}n}, {Tinker}, {Bundy},
  {Xhakaj}, \& {Wyithe}}]{Oyarzun2024}
{Oyarz{\'u}n}, G.~A., {Tinker}, J.~L., {Bundy}, K., {Xhakaj}, E., \& {Wyithe},
  J. S.~B. 2024,
  \href{http://dx.doi.org/10.3847/1538-4357/ad6de1}{\color{blue}\apj},
  \href{https://ui.adsabs.harvard.edu/abs/2024ApJ...974...29O}{974, 29}

\bibitem[{{Pacifici} {et~al.}(2016){Pacifici}, {Kassin}, {Weiner}, {Holden},
  {Gardner}, {Faber}, {Ferguson}, {Koo}, {Primack}, {Bell}, {Dekel}, {Gawiser},
  {Giavalisco}, {Rafelski}, {Simons}, {Barro}, {Croton}, {Dav{\'e}}, {Fontana},
  {Grogin}, {Koekemoer}, {Lee}, {Salmon}, {Somerville}, \&
  {Behroozi}}]{Pacifici2016}
{Pacifici}, C., {Kassin}, S.~A., {Weiner}, B.~J., {et~al.} 2016,
  \href{http://dx.doi.org/10.3847/0004-637X/832/1/79}{\color{blue}\apj},
  \href{https://ui.adsabs.harvard.edu/abs/2016ApJ...832...79P}{832, 79}

\bibitem[{{Pahwa} \& {Paranjape}(2017)}]{Pahwa2017}
{Pahwa}, I. \& {Paranjape}, A. 2017,
  \href{http://dx.doi.org/10.1093/mnras/stx1325}{\color{blue}\mnras},
  \href{https://ui.adsabs.harvard.edu/abs/2017MNRAS.470.1298P}{470, 1298}

\bibitem[{{Paquereau} {et~al.}(2025){Paquereau}, {Laigle}, {McCracken},
  {Shuntov}, {Ilbert}, {Akins}, {Allen}, {Arango-Togo}, {Berman},
  {B{\'e}thermin}, {Casey}, {McCleary}, {Dubois}, {Drakos}, {Faisst}, {Franco},
  {Harish}, {Jespersen}, {Kartaltepe}, {Koekemoer}, {Kokorev}, {Lambrides},
  {Larson}, {Liu}, {Le Borgne}, {Lewis}, {McKinney}, {Mercier}, {Rhodes},
  {Robertson}, {Toft}, {Trebitsch}, {Tresse}, \& {Weaver}}]{Paquereau2025}
{Paquereau}, L., {Laigle}, C., {McCracken}, H.~J., {et~al.} 2025,
  \href{http://dx.doi.org/10.1051/0004-6361/202553828}{\color{blue}\aap},
  \href{https://ui.adsabs.harvard.edu/abs/2025A&A...702A.163P}{702, A163}

\bibitem[{{Peng} {et~al.}(2015){Peng}, {Maiolino}, \& {Cochrane}}]{Peng2015}
{Peng}, Y., {Maiolino}, R., \& {Cochrane}, R. 2015,
  \href{http://dx.doi.org/10.1038/nature14439}{\color{blue}\nat},
  \href{https://ui.adsabs.harvard.edu/abs/2015Natur.521..192P}{521, 192}

\bibitem[{{Peng} {et~al.}(2010){Peng}, {Lilly}, {Kova{\v{c}}}, {Bolzonella},
  {Pozzetti}, {Renzini}, {Zamorani}, {Ilbert}, {Knobel}, {Iovino}, {Maier},
  {Cucciati}, {Tasca}, {Carollo}, {Silverman}, {Kampczyk}, {de Ravel},
  {Sanders}, {Scoville}, {Contini}, {Mainieri}, {Scodeggio}, {Kneib}, {Le
  F{\`e}vre}, {Bardelli}, {Bongiorno}, {Caputi}, {Coppa}, {de la Torre},
  {Franzetti}, {Garilli}, {Lamareille}, {Le Borgne}, {Le Brun}, {Mignoli},
  {Perez Montero}, {Pello}, {Ricciardelli}, {Tanaka}, {Tresse}, {Vergani},
  {Welikala}, {Zucca}, {Oesch}, {Abbas}, {Barnes}, {Bordoloi}, {Bottini},
  {Cappi}, {Cassata}, {Cimatti}, {Fumana}, {Hasinger}, {Koekemoer},
  {Leauthaud}, {Maccagni}, {Marinoni}, {McCracken}, {Memeo}, {Meneux}, {Nair},
  {Porciani}, {Presotto}, \& {Scaramella}}]{Peng2010}
{Peng}, Y.-j., {Lilly}, S.~J., {Kova{\v{c}}}, K., {et~al.} 2010,
  \href{http://dx.doi.org/10.1088/0004-637X/721/1/193}{\color{blue}\apj},
  \href{https://ui.adsabs.harvard.edu/abs/2010ApJ...721..193P}{721, 193}

\bibitem[{{Planck Collaboration} {et~al.}(2020){Planck Collaboration},
  {Aghanim}, {Akrami}, {Ashdown}, {Aumont}, {Baccigalupi}, {Ballardini},
  {Banday}, {Barreiro}, {Bartolo}, {Basak}, {Battye}, {Benabed}, {Bernard},
  {Bersanelli}, {Bielewicz}, {Bock}, {Bond}, {Borrill}, {Bouchet}, {Boulanger},
  {Bucher}, {Burigana}, {Butler}, {Calabrese}, {Cardoso}, {Carron},
  {Challinor}, {Chiang}, {Chluba}, {Colombo}, {Combet}, {Contreras}, {Crill},
  {Cuttaia}, {de Bernardis}, {de Zotti}, {Delabrouille}, {Delouis}, {Di
  Valentino}, {Diego}, {Dor{\'e}}, {Douspis}, {Ducout}, {Dupac}, {Dusini},
  {Efstathiou}, {Elsner}, {En{\ss}lin}, {Eriksen}, {Fantaye}, {Farhang},
  {Fergusson}, {Fernandez-Cobos}, {Finelli}, {Forastieri}, {Frailis},
  {Fraisse}, {Franceschi}, {Frolov}, {Galeotta}, {Galli}, {Ganga},
  {G{\'e}nova-Santos}, {Gerbino}, {Ghosh}, {Gonz{\'a}lez-Nuevo}, {G{\'o}rski},
  {Gratton}, {Gruppuso}, {Gudmundsson}, {Hamann}, {Handley}, {Hansen},
  {Herranz}, {Hildebrandt}, {Hivon}, {Huang}, {Jaffe}, {Jones}, {Karakci},
  {Keih{\"a}nen}, {Keskitalo}, {Kiiveri}, {Kim}, {Kisner}, {Knox},
  {Krachmalnicoff}, {Kunz}, {Kurki-Suonio}, {Lagache}, {Lamarre}, {Lasenby},
  {Lattanzi}, {Lawrence}, {Le Jeune}, {Lemos}, {Lesgourgues}, {Levrier},
  {Lewis}, {Liguori}, {Lilje}, {Lilley}, {Lindholm}, {L{\'o}pez-Caniego},
  {Lubin}, {Ma}, {Mac{\'\i}as-P{\'e}rez}, {Maggio}, {Maino}, {Mandolesi},
  {Mangilli}, {Marcos-Caballero}, {Maris}, {Martin}, {Martinelli},
  {Mart{\'\i}nez-Gonz{\'a}lez}, {Matarrese}, {Mauri}, {McEwen}, {Meinhold},
  {Melchiorri}, {Mennella}, {Migliaccio}, {Millea}, {Mitra},
  {Miville-Desch{\^e}nes}, {Molinari}, {Montier}, {Morgante}, {Moss}, {Natoli},
  {N{\o}rgaard-Nielsen}, {Pagano}, {Paoletti}, {Partridge}, {Patanchon},
  {Peiris}, {Perrotta}, {Pettorino}, {Piacentini}, {Polastri}, {Polenta},
  {Puget}, {Rachen}, {Reinecke}, {Remazeilles}, {Renzi}, {Rocha}, {Rosset},
  {Roudier}, {Rubi{\~n}o-Mart{\'\i}n}, {Ruiz-Granados}, {Salvati}, {Sandri},
  {Savelainen}, {Scott}, {Shellard}, {Sirignano}, {Sirri}, {Spencer},
  {Sunyaev}, {Suur-Uski}, {Tauber}, {Tavagnacco}, {Tenti}, {Toffolatti},
  {Tomasi}, {Trombetti}, {Valenziano}, {Valiviita}, {Van Tent}, {Vibert},
  {Vielva}, {Villa}, {Vittorio}, {Wandelt}, {Wehus}, {White}, {White},
  {Zacchei}, \& {Zonca}}]{Planck18}
{Planck Collaboration}, {Aghanim}, N., {Akrami}, Y., {et~al.} 2020,
  \href{http://dx.doi.org/10.1051/0004-6361/201833910}{\color{blue}\aap},
  \href{https://ui.adsabs.harvard.edu/abs/2020A&A...641A...6P}{641, A6}

\bibitem[{{Pozzetti} {et~al.}(2010){Pozzetti}, {Bolzonella}, {Zucca},
  {Zamorani}, {Lilly}, {Renzini}, {Moresco}, {Mignoli}, {Cassata}, {Tasca},
  {Lamareille}, {Maier}, {Meneux}, {Halliday}, {Oesch}, {Vergani}, {Caputi},
  {Kova{\v{c}}}, {Cimatti}, {Cucciati}, {Iovino}, {Peng}, {Carollo}, {Contini},
  {Kneib}, {Le F{\'e}vre}, {Mainieri}, {Scodeggio}, {Bardelli}, {Bongiorno},
  {Coppa}, {de la Torre}, {de Ravel}, {Franzetti}, {Garilli}, {Kampczyk},
  {Knobel}, {Le Borgne}, {Le Brun}, {Pell{\`o}}, {Perez Montero},
  {Ricciardelli}, {Silverman}, {Tanaka}, {Tresse}, {Abbas}, {Bottini}, {Cappi},
  {Guzzo}, {Koekemoer}, {Leauthaud}, {Maccagni}, {Marinoni}, {McCracken},
  {Memeo}, {Porciani}, {Scaramella}, {Scarlata}, \& {Scoville}}]{Pozzetti10}
{Pozzetti}, L., {Bolzonella}, M., {Zucca}, E., {et~al.} 2010,
  \href{http://dx.doi.org/10.1051/0004-6361/200913020}{\color{blue}\aap},
  \href{https://ui.adsabs.harvard.edu/abs/2010A&A...523A..13P}{523, A13}

\bibitem[{{Roche} \& {Eales}(1999)}]{RocheEales99}
{Roche}, N. \& {Eales}, S.~A. 1999,
  \href{http://dx.doi.org/10.1046/j.1365-8711.1999.02652.x}{\color{blue}\mnras},
  \href{https://ui.adsabs.harvard.edu/abs/1999MNRAS.307..703R}{307, 703}

\bibitem[{{Rodr{\'\i}guez-Puebla} {et~al.}(2015){Rodr{\'\i}guez-Puebla},
  {Avila-Reese}, {Yang}, {Foucaud}, {Drory}, \& {Jing}}]{RodriguezPuebla2015}
{Rodr{\'\i}guez-Puebla}, A., {Avila-Reese}, V., {Yang}, X., {et~al.} 2015,
  \href{http://dx.doi.org/10.1088/0004-637X/799/2/130}{\color{blue}\apj},
  \href{https://ui.adsabs.harvard.edu/abs/2015ApJ...799..130R}{799, 130}

\bibitem[{{Sato} {et~al.}(2014){Sato}, {Sawicki}, \& {Arcila-Osejo}}]{Sato2014}
{Sato}, T., {Sawicki}, M., \& {Arcila-Osejo}, L. 2014,
  \href{http://dx.doi.org/10.1093/mnras/stu1356}{\color{blue}\mnras},
  \href{https://ui.adsabs.harvard.edu/abs/2014MNRAS.443.2661S}{443, 2661}

\bibitem[{{Schechter}(1976)}]{Schechter1976}
{Schechter}, P. 1976,
  \href{http://dx.doi.org/10.1086/154079}{\color{blue}\apj},
  \href{https://ui.adsabs.harvard.edu/abs/1976ApJ...203..297S}{203, 297}

\bibitem[{{Scognamiglio} {et~al.}(2026){Scognamiglio}, {Leroy}, {Harvey},
  {Massey}, {Rhodes}, {Akins}, {Brinch}, {Berman}, {Casey}, {Drakos}, {Faisst},
  {Franco}, {Fung}, {Gozaliasl}, {He}, {Hatamnia}, {Huff}, {Hogg}, {Ilbert},
  {Kartaltepe}, {Koekemoer}, {Jin}, {Lambrides}, {Leauthaud}, {Lentz}, {Liu},
  {Mahler}, {Maraston}, {Martin}, {McCleary}, {Nightingale}, {Mobasher},
  {Paquereau}, {Pires}, {Robertson}, {Sanders}, {Scarlata}, {Shuntov}, {Toni},
  {von Wietersheim-Kramsta}, \& {Weaver}}]{Scognamiglio2026}
{Scognamiglio}, D., {Leroy}, G., {Harvey}, D., {et~al.} 2026,
  \href{http://dx.doi.org/10.1038/s41550-025-02763-9}{\color{blue}Nature
  Astronomy}
  \href{https://ui.adsabs.harvard.edu/abs/2026NatAs.tmp...25S}{[\eprint[arXiv]{2601.17239}]}

\bibitem[{{Shi} {et~al.}(2024){Shi}, {Malavasi}, {Toshikawa}, \&
  {Zheng}}]{Shi2024}
{Shi}, K., {Malavasi}, N., {Toshikawa}, J., \& {Zheng}, X. 2024,
  \href{http://dx.doi.org/10.3847/1538-4357/ad11d7}{\color{blue}\apj},
  \href{https://ui.adsabs.harvard.edu/abs/2024ApJ...961...39S}{961, 39}

\bibitem[{{Shuntov} {et~al.}(2025){Shuntov}, {Akins}, {Paquereau}, {Casey},
  {Ilbert}, {Arango-Toro}, {McCracken}, {Franco}, {Harish}, {Kartaltepe},
  {Koekemoer}, {Yang}, {Huertas-Company}, {Berman}, {McCleary}, {Toft},
  {Gavazzi}, {Achenbach}, {Bertin}, {Brinch}, {Champagne}, {Chartab}, {Drakos},
  {Egami}, {Endsley}, {Faisst}, {Fan}, {Flayhart}, {Hartley}, {Hatamnia},
  {Gozaliasl}, {Gentile}, {Jermann}, {Jin}, {Kakiichi}, {Khostovan},
  {K{\"u}mmel}, {Laigle}, {Laishram}, {Lambrides}, {Liu}, {Lyu}, {Magdis},
  {Mobasher}, {Moutard}, {Renzini}, {Rich}, {Sanders}, {Sattari}, {Robertson},
  {Schefer}, {Scognamiglio}, {Scoville}, {Silverman}, {Taamoli},
  {Trakhtenbrot}, {Valentino}, {Wang}, {Weaver}, \& {Yang}}]{Shuntov2025d}
{Shuntov}, M., {Akins}, H.~B., {Paquereau}, L., {et~al.} 2025,
  \href{http://dx.doi.org/10.1051/0004-6361/202555799}{\color{blue}\aap},
  \href{https://ui.adsabs.harvard.edu/abs/2025A&A...704A.339S}{704, A339}

\bibitem[{{Shuntov} {et~al.}(2026){Shuntov}, {Ilbert}, {Lagos}, {Toft},
  {Valentino}, {Mercier}, {Akins}, {Binh}, {Brinch}, {Casey}, {Franco},
  {Gentile}, {Gozaliasl}, {Haghjoo}, {Harish}, {Hirschmann}, {Huertas-Company},
  {Jin}, {Kartaltepe}, {Koekemoer}, {Laigle}, {Lewis}, {Magdis}, {Joy
  McCracken}, {Mobasher}, {Moutard}, {Oesch}, {Paquereau}, {Renzini}, {Rich},
  {Sanders}, {Toni}, {Tresse}, {Weibel}, {Weaver}, \& {Yang}}]{Shuntov2026b}
{Shuntov}, M., {Ilbert}, O., {Lagos}, C. d.~P., {et~al.} 2026,
  \href{http://dx.doi.org/10.1051/0004-6361/202558022}{\color{blue}\aap},
  \href{https://ui.adsabs.harvard.edu/abs/2026A&A...707A.391S}{707, A391}

\bibitem[{{Singh} {et~al.}(2025){Singh}, {Guaita}, {Hibon}, {H{\"a}u{\ss}ler},
  {Lee}, {Ramakrishnan}, {Kumar}, {Padilla}, {Firestone}, {Song}, {Artale},
  {Hwang}, {Iribarren}, {Gronwall}, {Gawiser}, {Nantais}, {Valdes}, {Park}, \&
  {Yang}}]{Singh2025}
{Singh}, A., {Guaita}, L., {Hibon}, P., {et~al.} 2025,
  \href{http://dx.doi.org/10.1051/0004-6361/202452406}{\color{blue}\aap},
  \href{https://ui.adsabs.harvard.edu/abs/2025A&A...700A..68S}{700, A68}

\bibitem[{{Song} {et~al.}(2021){Song}, {Laigle}, {Hwang}, {Devriendt},
  {Dubois}, {Kraljic}, {Pichon}, {Slyz}, \& {Smith}}]{Song2021}
{Song}, H., {Laigle}, C., {Hwang}, H.~S., {et~al.} 2021,
  \href{http://dx.doi.org/10.1093/mnras/staa3981}{\color{blue}\mnras},
  \href{https://ui.adsabs.harvard.edu/abs/2021MNRAS.501.4635S}{501, 4635}

\bibitem[{{Sun} {et~al.}(2018){Sun}, {Guo}, {Wang}, {Lacey}, {Wang}, {Gao}, \&
  {Pan}}]{Sun2018}
{Sun}, S., {Guo}, Q., {Wang}, L., {et~al.} 2018,
  \href{http://dx.doi.org/10.1093/mnras/sty832}{\color{blue}\mnras},
  \href{https://ui.adsabs.harvard.edu/abs/2018MNRAS.477.3136S}{477, 3136}

\bibitem[{{Taamoli} {et~al.}(2024){Taamoli}, {Mobasher}, {Chartab}, {Darvish},
  {Weaver}, {Hemmati}, {Casey}, {Sattari}, {Brammer}, {Capak}, {Ilbert},
  {Kartaltepe}, {McCracken}, {Moneti}, {Sanders}, {Scoville}, {Steinhardt}, \&
  {Toft}}]{Taamoli2024}
{Taamoli}, S., {Mobasher}, B., {Chartab}, N., {et~al.} 2024,
  \href{http://dx.doi.org/10.3847/1538-4357/ad32c5}{\color{blue}\apj},
  \href{https://ui.adsabs.harvard.edu/abs/2024ApJ...966...18T}{966, 18}

\bibitem[{{Tacchella} {et~al.}(2019){Tacchella}, {Diemer}, {Hernquist},
  {Genel}, {Marinacci}, {Nelson}, {Pillepich}, {Rodriguez-Gomez}, {Sales},
  {Springel}, \& {Vogelsberger}}]{Tacchella2019}
{Tacchella}, S., {Diemer}, B., {Hernquist}, L., {et~al.} 2019,
  \href{http://dx.doi.org/10.1093/mnras/stz1657}{\color{blue}\mnras},
  \href{https://ui.adsabs.harvard.edu/abs/2019MNRAS.487.5416T}{487, 5416}

\bibitem[{{Tinker} {et~al.}(2018){Tinker}, {Hahn}, {Mao}, {Wetzel}, \&
  {Conroy}}]{Tinker2018a}
{Tinker}, J.~L., {Hahn}, C., {Mao}, Y.-Y., {Wetzel}, A.~R., \& {Conroy}, C.
  2018, \href{http://dx.doi.org/10.1093/mnras/sty666}{\color{blue}\mnras},
  \href{https://ui.adsabs.harvard.edu/abs/2018MNRAS.477..935T}{477, 935}

\bibitem[{{Tinker} {et~al.}(2013){Tinker}, {Leauthaud}, {Bundy}, {George},
  {Behroozi}, {Massey}, {Rhodes}, \& {Wechsler}}]{Tinker2013}
{Tinker}, J.~L., {Leauthaud}, A., {Bundy}, K., {et~al.} 2013,
  \href{http://dx.doi.org/10.1088/0004-637X/778/2/93}{\color{blue}\apj},
  \href{https://ui.adsabs.harvard.edu/abs/2013ApJ...778...93T}{778, 93}

\bibitem[{{Tojeiro} \& {Kraljic}(2025)}]{Tojeiro&Kraljic2025}
{Tojeiro}, R. \& {Kraljic}, K. 2025,
  \href{https://ui.adsabs.harvard.edu/abs/2025arXiv250321759T}{\href{http://dx.doi.org/10.48550/arXiv.2503.21759}{\color{blue}arXiv
  e-prints}, arXiv:2503.21759}

\bibitem[{{Toni} {et~al.}(2026{\natexlab{a}}){Toni}, {Castignani}, {Combes},
  {Salom{\'e}}, {Bongiovanni}, {Moscardini}, \& {Maturi}}]{Toni2026}
{Toni}, G., {Castignani}, G., {Combes}, F., {et~al.} 2026{\natexlab{a}},
  \href{https://ui.adsabs.harvard.edu/abs/2026arXiv260521592T}{\href{http://dx.doi.org/10.48550/arXiv.2605.21592}{\color{blue}arXiv
  e-prints}, arXiv:2605.21592}

\bibitem[{{Toni} {et~al.}(2025){Toni}, {Gozaliasl}, {Maturi}, {Moscardini},
  {Finoguenov}, {Castignani}, {Gentile}, {Virolainen}, {Casey}, {Kartaltepe},
  {Akins}, {Allen}, {Arango-Toro}, {Babul}, {Brinch}, {Drakos}, {Faisst},
  {Franco}, {Griffiths}, {Harish}, {Hasinger}, {Ilbert}, {Jin}, {Khostovan},
  {Koekemoer}, {Korpi-Lagg}, {Larson}, {Lertprasertpong}, {Liu}, {Magdis},
  {Massey}, {McCracken}, {McKinney}, {Paquereau}, {Rhodes}, {Robertson},
  {Sargent}, {Shuntov}, {Tanaka}, {Taamoli}, {Tempel}, {Toft}, {Vardoulaki}, \&
  {Yang}}]{Toni2025}
{Toni}, G., {Gozaliasl}, G., {Maturi}, M., {et~al.} 2025,
  \href{http://dx.doi.org/10.1051/0004-6361/202553759}{\color{blue}\aap},
  \href{https://ui.adsabs.harvard.edu/abs/2025A&A...697A.197T}{697, A197}

\bibitem[{{Toni} {et~al.}(2026{\natexlab{b}}){Toni}, {Maturi}, {Castignani},
  {Moscardini}, {Gozaliasl}, {Finoguenov}, {Taamoli}, {Gentile}, {Akins},
  {Arango-Toro}, {Casey}, {Drakos}, {Faisst}, {Flayhart}, {Franco}, {Hadi},
  {Haghjoo}, {Harish}, {Hatamnia}, {Ilbert}, {Jin}, {Kartaltepe}, {Khostovan},
  {Koekemoer}, {Leroy}, {Magdis}, {McCracken}, {McKinney}, {Paquereau},
  {Rhodes}, {Rich}, {Robertson}, {Samir}, {Scognamiglio}, {Shamyati},
  {Shuntov}, \& {Zavala}}]{Toni2026a}
{Toni}, G., {Maturi}, M., {Castignani}, G., {et~al.} 2026{\natexlab{b}},
  \href{http://dx.doi.org/10.1051/0004-6361/202557183}{\color{blue}\aap},
  \href{https://ui.adsabs.harvard.edu/abs/2026A&A...707A..87T}{707, A87}

\bibitem[{{Toni} {et~al.}(2024){Toni}, {Maturi}, {Finoguenov}, {Moscardini}, \&
  {Castignani}}]{Toni2024}
{Toni}, G., {Maturi}, M., {Finoguenov}, A., {Moscardini}, L., \& {Castignani},
  G. 2024,
  \href{http://dx.doi.org/10.1051/0004-6361/202348832}{\color{blue}\aap},
  \href{https://ui.adsabs.harvard.edu/abs/2024A&A...687A..56T}{687, A56}

\bibitem[{{Treyer} {et~al.}(2018){Treyer}, {Kraljic}, {Arnouts}, {de la Torre},
  {Pichon}, {Dubois}, {Vibert}, {Milliard}, {Laigle}, {Seibert}, {Brown},
  {Grootes}, {Wright}, {Liske}, {Lara-Lopez}, \& {Bland-Hawthorn}}]{Treyer2018}
{Treyer}, M., {Kraljic}, K., {Arnouts}, S., {et~al.} 2018,
  \href{http://dx.doi.org/10.1093/mnras/sty769}{\color{blue}\mnras},
  \href{https://ui.adsabs.harvard.edu/abs/2018MNRAS.477.2684T}{477, 2684}

\bibitem[{{Valentino} {et~al.}(2020){Valentino}, {Tanaka}, {Davidzon}, {Toft},
  {G{\'o}mez-Guijarro}, {Stockmann}, {Onodera}, {Brammer}, {Ceverino},
  {Faisst}, {Gallazzi}, {Hayward}, {Ilbert}, {Kubo}, {Magdis}, {Selsing},
  {Shimakawa}, {Sparre}, {Steinhardt}, {Yabe}, \& {Zabl}}]{Valentino2020}
{Valentino}, F., {Tanaka}, M., {Davidzon}, I., {et~al.} 2020,
  \href{http://dx.doi.org/10.3847/1538-4357/ab64dc}{\color{blue}\apj},
  \href{https://ui.adsabs.harvard.edu/abs/2020ApJ...889...93V}{889, 93}

\bibitem[{{van den Bosch} {et~al.}(2003){van den Bosch}, {Yang}, \&
  {Mo}}]{vandenBosch2003}
{van den Bosch}, F.~C., {Yang}, X., \& {Mo}, H.~J. 2003,
  \href{http://dx.doi.org/10.1046/j.1365-8711.2003.06335.x}{\color{blue}\mnras},
  \href{https://ui.adsabs.harvard.edu/abs/2003MNRAS.340..771V}{340, 771}

\bibitem[{{Wang} {et~al.}(2022){Wang}, {Mao}, {Zentner}, {Guo}, {Lange}, {van
  den Bosch}, \& {Mezini}}]{Wang2022}
{Wang}, K., {Mao}, Y.-Y., {Zentner}, A.~R., {et~al.} 2022,
  \href{http://dx.doi.org/10.1093/mnras/stac2465}{\color{blue}\mnras},
  \href{https://ui.adsabs.harvard.edu/abs/2022MNRAS.516.4003W}{516, 4003}

\bibitem[{{Wang} {et~al.}(2026){Wang}, {Schaye}, {Ben{\'\i}tez-Llambay},
  {Chaikin}, {Frenk}, {Hu{\v{s}}ko}, {McGibbon}, {Ploeckinger}, {Richings},
  {Schaller}, \& {Trayford}}]{Wang2026}
{Wang}, K., {Schaye}, J., {Ben{\'\i}tez-Llambay}, A., {et~al.} 2026,
  \href{http://dx.doi.org/10.1093/mnras/stag110}{\color{blue}\mnras},
  \href{https://ui.adsabs.harvard.edu/abs/2026MNRAS.546ag110W}{546, stag110}

\bibitem[{{Wang} \& {White}(2012)}]{Wang&White2012}
{Wang}, W. \& {White}, S. D.~M. 2012,
  \href{http://dx.doi.org/10.1111/j.1365-2966.2012.21256.x}{\color{blue}\mnras},
  \href{https://ui.adsabs.harvard.edu/abs/2012MNRAS.424.2574W}{424, 2574}

\bibitem[{{Watson} {et~al.}(2015){Watson}, {Hearin}, {Berlind}, {Becker},
  {Behroozi}, {Skibba}, {Reyes}, {Zentner}, \& {van den Bosch}}]{Watson2015}
{Watson}, D.~F., {Hearin}, A.~P., {Berlind}, A.~A., {et~al.} 2015,
  \href{http://dx.doi.org/10.1093/mnras/stu2065}{\color{blue}\mnras},
  \href{https://ui.adsabs.harvard.edu/abs/2015MNRAS.446..651W}{446, 651}

\bibitem[{{Wechsler} \& {Tinker}(2018)}]{Wechsler&Tinker2018}
{Wechsler}, R.~H. \& {Tinker}, J.~L. 2018,
  \href{http://dx.doi.org/10.1146/annurev-astro-081817-051756}{\color{blue}\araa},
  \href{https://ui.adsabs.harvard.edu/abs/2018ARA&A..56..435W}{56, 435}

\bibitem[{{Weinmann} {et~al.}(2006){Weinmann}, {van den Bosch}, {Yang}, \&
  {Mo}}]{Weinmann2006}
{Weinmann}, S.~M., {van den Bosch}, F.~C., {Yang}, X., \& {Mo}, H.~J. 2006,
  \href{http://dx.doi.org/10.1111/j.1365-2966.2005.09865.x}{\color{blue}\mnras},
  \href{https://ui.adsabs.harvard.edu/abs/2006MNRAS.366....2W}{366, 2}

\bibitem[{{Wetzel} {et~al.}(2012){Wetzel}, {Tinker}, \& {Conroy}}]{Wetzel2012}
{Wetzel}, A.~R., {Tinker}, J.~L., \& {Conroy}, C. 2012,
  \href{http://dx.doi.org/10.1111/j.1365-2966.2012.21188.x}{\color{blue}\mnras},
  \href{https://ui.adsabs.harvard.edu/abs/2012MNRAS.424..232W}{424, 232}

\bibitem[{{White} \& {Rees}(1978)}]{White&Rees1978}
{White}, S.~D.~M. \& {Rees}, M.~J. 1978,
  \href{http://dx.doi.org/10.1093/mnras/183.3.341}{\color{blue}\mnras},
  \href{https://ui.adsabs.harvard.edu/abs/1978MNRAS.183..341W}{183, 341}

\bibitem[{{Williams} {et~al.}(2009){Williams}, {Quadri}, {Franx}, {van Dokkum},
  \& {Labb{\'e}}}]{Williams2009}
{Williams}, R.~J., {Quadri}, R.~F., {Franx}, M., {van Dokkum}, P., \&
  {Labb{\'e}}, I. 2009,
  \href{http://dx.doi.org/10.1088/0004-637X/691/2/1879}{\color{blue}\apj},
  \href{https://ui.adsabs.harvard.edu/abs/2009ApJ...691.1879W}{691, 1879}

\bibitem[{{Zaidi} {et~al.}(2026){Zaidi}, {Wake}, {Marchesini}, {Iyer},
  {Muzzin}, {Papovich}, {Antwi-Danso}, {Glazebrook}, \&
  {Labb{\'e}}}]{Zaidi2024}
{Zaidi}, K., {Wake}, D.~A., {Marchesini}, D., {et~al.} 2026,
  \href{http://dx.doi.org/10.3847/1538-4357/ae6b7d}{\color{blue}\apj},
  \href{https://ui.adsabs.harvard.edu/abs/2026ApJ..1004...96Z}{1004, 96}

\bibitem[{{Zehavi} {et~al.}(2011){Zehavi}, {Zheng}, {Weinberg}, {Blanton},
  {Bahcall}, {Berlind}, {Brinkmann}, {Frieman}, {Gunn}, {Lupton}, {Nichol},
  {Percival}, {Schneider}, {Skibba}, {Strauss}, {Tegmark}, \&
  {York}}]{Zehavi2011}
{Zehavi}, I., {Zheng}, Z., {Weinberg}, D.~H., {et~al.} 2011,
  \href{http://dx.doi.org/10.1088/0004-637X/736/1/59}{\color{blue}\apj},
  \href{https://ui.adsabs.harvard.edu/abs/2011ApJ...736...59Z}{736, 59}

\bibitem[{{Zehavi} {et~al.}(2005){Zehavi}, {Zheng}, {Weinberg}, {Frieman},
  {Berlind}, {Blanton}, {Scoccimarro}, {Sheth}, {Strauss}, {Kayo}, {Suto},
  {Fukugita}, {Nakamura}, {Bahcall}, {Brinkmann}, {Gunn}, {Hennessy},
  {Ivezi{\'c}}, {Knapp}, {Loveday}, {Meiksin}, {Schlegel}, {Schneider},
  {Szapudi}, {Tegmark}, {Vogeley}, {York}, \& {SDSS
  Collaboration}}]{Zehavi2005}
{Zehavi}, I., {Zheng}, Z., {Weinberg}, D.~H., {et~al.} 2005,
  \href{http://dx.doi.org/10.1086/431891}{\color{blue}\apj},
  \href{https://ui.adsabs.harvard.edu/abs/2005ApJ...630....1Z}{630, 1}

\bibitem[{{Zentner} {et~al.}(2019){Zentner}, {Hearin}, {van den Bosch},
  {Lange}, \& {Villarreal}}]{Zentner2019}
{Zentner}, A.~R., {Hearin}, A., {van den Bosch}, F.~C., {Lange}, J.~U., \&
  {Villarreal}, A.~S. 2019,
  \href{http://dx.doi.org/10.1093/mnras/stz470}{\color{blue}\mnras},
  \href{https://ui.adsabs.harvard.edu/abs/2019MNRAS.485.1196Z}{485, 1196}

\bibitem[{{Zentner} {et~al.}(2014){Zentner}, {Hearin}, \& {van den
  Bosch}}]{Zentner2014}
{Zentner}, A.~R., {Hearin}, A.~P., \& {van den Bosch}, F.~C. 2014,
  \href{http://dx.doi.org/10.1093/mnras/stu1383}{\color{blue}\mnras},
  \href{https://ui.adsabs.harvard.edu/abs/2014MNRAS.443.3044Z}{443, 3044}

\bibitem[{{Zu} \& {Mandelbaum}(2016)}]{Zu&Mandelbaum2016}
{Zu}, Y. \& {Mandelbaum}, R. 2016,
  \href{http://dx.doi.org/10.1093/mnras/stw221}{\color{blue}\mnras},
  \href{https://ui.adsabs.harvard.edu/abs/2016MNRAS.457.4360Z}{457, 4360}

\bibitem[{{Zu} \& {Mandelbaum}(2018)}]{Zu&Mandelbaum2018}
{Zu}, Y. \& {Mandelbaum}, R. 2018,
  \href{http://dx.doi.org/10.1093/mnras/sty279}{\color{blue}\mnras},
  \href{https://ui.adsabs.harvard.edu/abs/2018MNRAS.476.1637Z}{476, 1637}

\end{thebibliography}

\begin{appendix}

\twocolumn

\section{Quiescent/star-forming classification} \label{app:class}

Various methods exist to classify galaxies into quiescent or star-forming populations, typically by employing either rest-frame colour-colour diagrams \citep[\eg $UVJ$,][]{Williams2009}, or exploiting SFR or sSFR thresholds \citep[\eg][]{Labbe2005}. We compared our evolving sSFR classification with alternatives, like a $NUVrJ$ selection \citep{Arnouts2007,Ilbert2013} or a 2$\sigma$ distance to the main sequence (noted $\Delta$MS here; \eg \citealt{Fang2018, Bluck2022}).

\begin{figure}[h!]
    \centering
    \includegraphics[width=1.0\columnwidth]{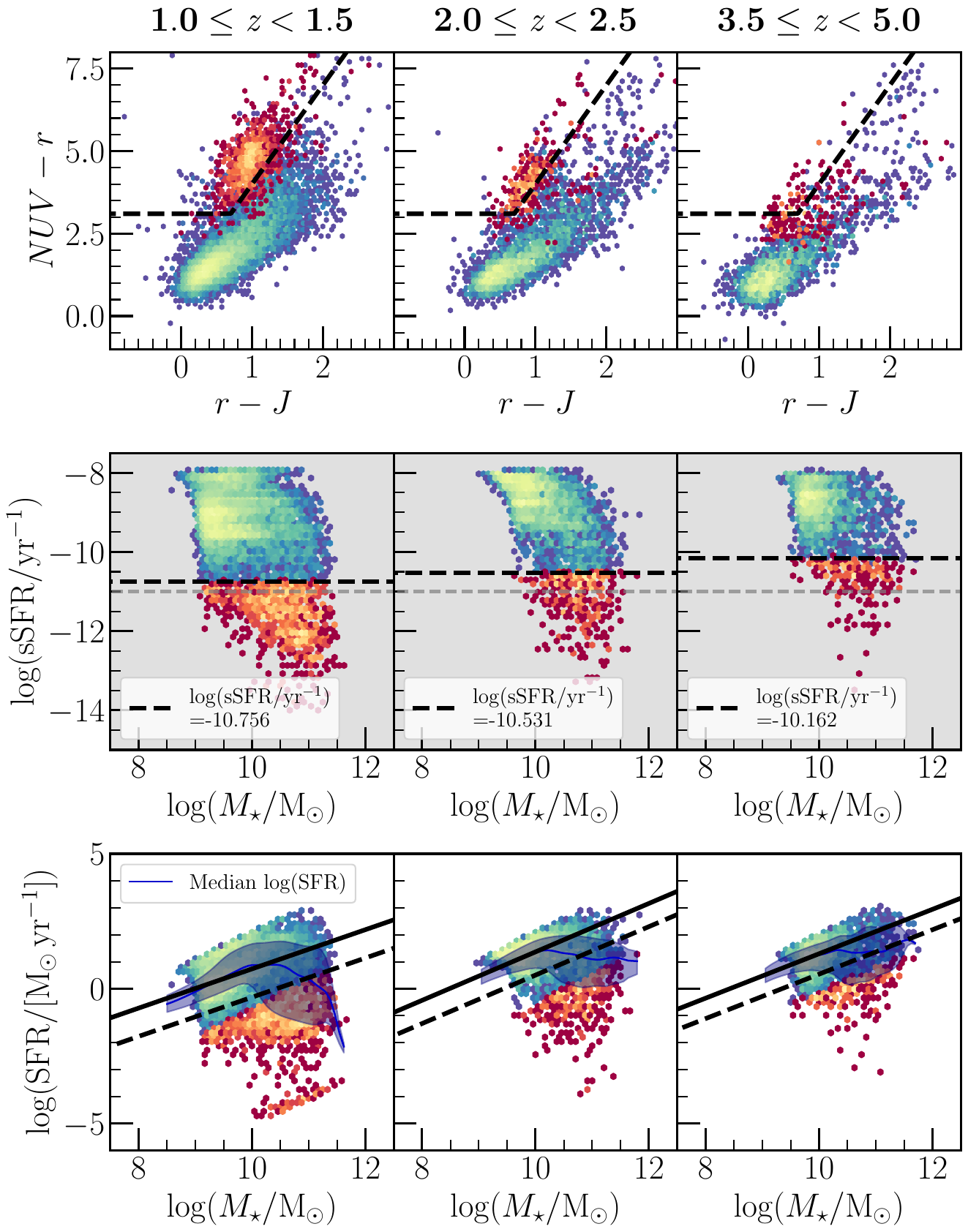}
    \caption{Quiescent/star-forming classification, selected by a threshold in sSFR evolving with redshift (middle row), in three redshift bins. The resulting QGs (in red) and SFGs (in blue) are also shown in a $NUVrJ$ diagram (top row, with the black line being the criterion) and in relation to the main sequence (bottom row, where the dashed line represents a $2\sigma$ distance to the linear fit of the MS shown with the black line).}
    \label{app-fig:sSFRclass_contamination}
\end{figure}

As shown in \reffig{app-fig:sSFRclass_contamination}, our identified QGs would also be identified as quiescent by these other methods. They also show a moderate clustering signal, in between the results of the other approaches. In contrast, these alternative selections show signs of contamination in different tracers: $\Delta$MS-selected QGs lie in the star-forming region of $NUVrJ$ at high-$z$ ($z \gtrsim 2$), or a large proportion of $NUVrJ$-selected QGs show high sSFR at low-$z$ ($z \lesssim 2$). 

A probabilistic alternative has been developed by \cite{Toni2026a}, who trained a machine-learning classifier on COSMOS-Web based on several traditional classification methods. This algorithm returns a ``quiescence probability'', that could in principle be used to weight clustering measurements; however, we leave this for future work.

\FloatBarrier

\section{Numbers of galaxies per bin} \label{app:numbers}

\begin{table}[h!]
    \centering
    \caption{Numbers of SFGs and QGs in each redshift and stellar mass threshold bins.}
    \fontsize{8.9pt}{8.9pt}\selectfont
    \renewcommand{\arraystretch}{1.51}
    \begin{tabular}{llllllllll}
    \hline \hline
    z bin               & $\log (M_\star^{\textrm{th}}/M_\odot)$ & $N_{\rm SFG}$ & $N_{\rm QG}$ \\ \hline
    $0.1 \le z < 0.6$   & 7.75 & 20628 & 2862 \\
                        & 8.5  & 6902 & 1760 \\
                        & 9.5  & 1577 & 928 \\
                        & 10.5 & 139 & 383 \\ \hline
    $0.6 \le z < 1.0$   & 8.0  & 42537 & 2744 \\
                        & 9.0  & 13935 & 2549 \\
                        & 10.0 & 3247 & 1794 \\
                        & 10.5 & 1091 & 1175 \\
                        & 11.0 & 95 & 360 \\ \hline
    $1.0 \le z < 1.5$   & 9.0  & 15001 & 1347 \\
                        & 10.0 & 2908 & 1147 \\
                        & 10.5 & 1066 & 798 \\
                        & 11.0 & 192 & 241 \\ \hline
    $1.5 \le z < 2.0$   & 9.0  & 16705 & 934 \\
                        & 10.0 & 3133 & 866 \\
                        & 10.75 & 706 & 424 \\ \hline
    $2.0 \le z < 2.5$   & 9.5  & 5806 & 376 \\
                        & 10.75 & 571 & 204 \\ \hline
    $2.5 \le z < 3.5$   & 9.5  & 7587 & 325 \\
                        & 10.5 & 1044 & 235 \\ \hline
    $3.5 \le z < 5.0$   & 9.5  & 4510 & 227 \\
                        & 10.5 & 448 & 133 \\ \hline  
    \end{tabular}
    \vspace{3mm}
    \label{app-tab:counts}
\end{table}
\FloatBarrier

\section{Quiescent fraction} \label{app:fQG}

\begin{figure}[h!]
    \centering
    \includegraphics[width=0.9\columnwidth]{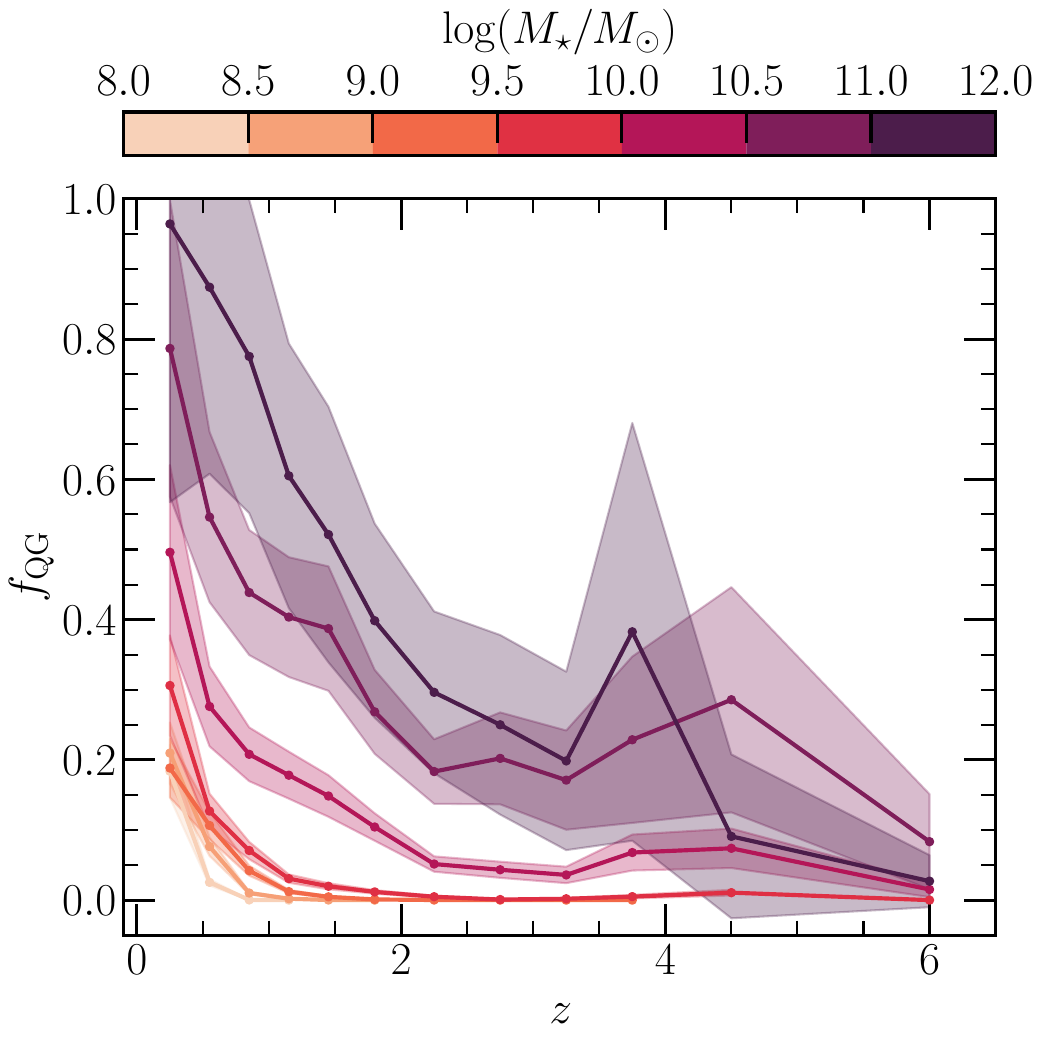}
    \caption{Fraction of quiescent galaxies (defined as $f_{\rm QG} = N_{\rm QG}/N_{\rm total}$) in stellar mass bins, as a function of redshift. Shaded areas show the $1\sigma$ uncertainty computed from Poisson errors and cosmic variance. }
    \label{app-fig:fQG}
\end{figure}

We show in \reffig{app-fig:fQG} the evolution with redshift of the fraction of QGs, within stellar mass bins. For each stellar mass bin, $f_{\rm QG}$ appears to mildly evolve from high-$z$ down to $z \sim 1 - 2$ (depending on $M_\star$), below which it increases exponentially. The peaks at $z \sim 3-4$ for the most massive QGs can be explained by the presence of a protocluster in the COSMOS-Web area, evidenced for instance by \cite{Toni2025}. 

This evolution is consistent with the conclusions of \refsubsec{subsec:envquenching}. Massive QGs have most likely been quenched by internal processes (such as AGN feedback), which can operate at any redshift and produce a gradual increase in $f_{\rm QG}$. For lower mass galaxies, however, $f_{\rm QG}$ rises sharply below a ``triggering'' redshift that depends on the mass bin, suggesting that environmental quenching becomes the dominant driver once structures such as groups and clusters have grown sufficiently to efficiently quench satellites and low-mass centrals. A similar conclusion was reached by \cite{Ji18} who found that the fraction of low-mass galaxies evolves more rapidly in the $z = 1 - 4$ range compared to massive ones. We refer the reader to \cite{Shuntov2026b} for a more complete analysis of quiescent fractions and stellar mass functions across cosmic time in COSMOS-Web.

\FloatBarrier

\section{Illustration of the halo mass matching procedure} \label{app:HMM}

\begin{figure}[h!]
    \centering
    \includegraphics[width=1.0\columnwidth]{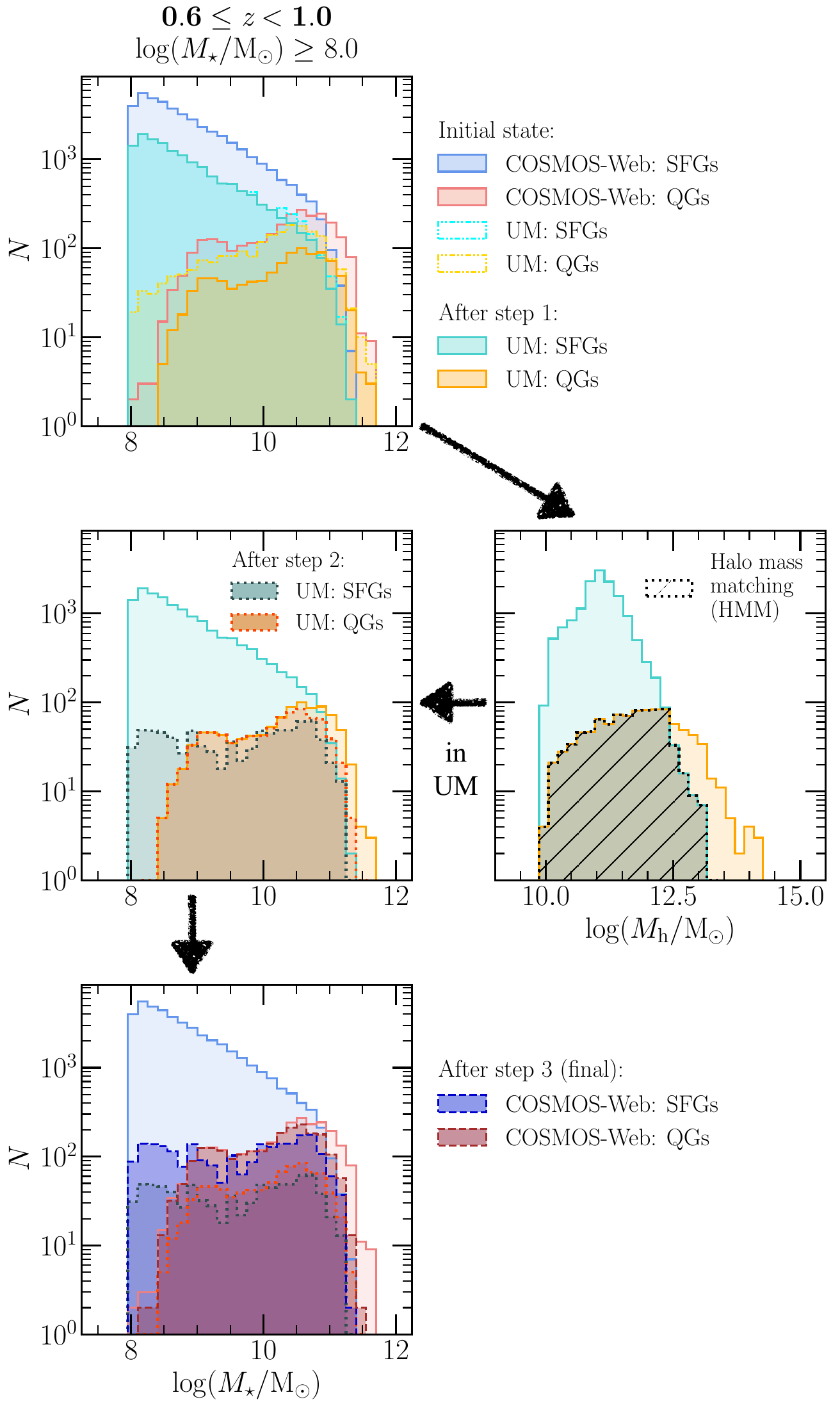}
    \caption{Halo mass-matching procedure between SFGs and QGs in COSMOS-Web at $0.6 \le z < 1$ using \textsc{UniverseMachine}. Each row represents a sequential step in the process: step 1 matches stellar mass distributions of \UM galaxies to COSMOS-Web ones; step 2 matches halo mass distributions between SFGs and QGs in \UM; and the last step matches the initial COSMOS-Web stellar mass distributions of SFGs and QGs to those resulting from step 2 in \UM.}
    \label{fig:halomassmatching}
\end{figure}

\clearpage
\onecolumn

\section{Auto-correlation of SFGs and QGs after halo mass matching} \label{app:autocorr_thresh_HMM}

\begin{figure*}[h!]
    \centering
    \includegraphics[height=0.93\textheight]{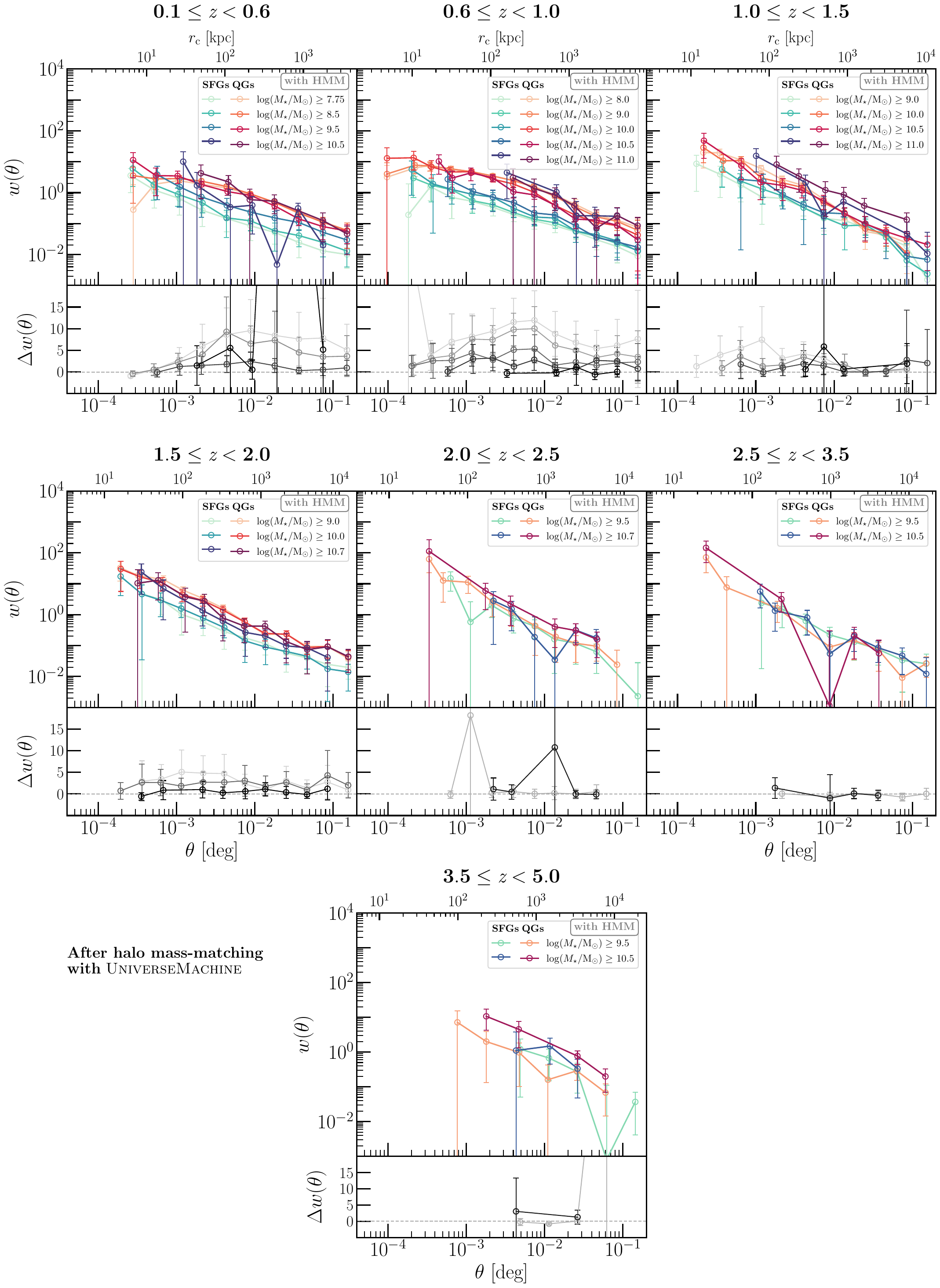}
    \caption{Similar to \reffig{fig:autocorr_thresh}, but after performing halo mass matching.}
    \label{app-fig:autocorr_thresh_HMM}
\end{figure*}

\begin{figure}[t!]
    \centering
    \includegraphics[width=0.95\textwidth]{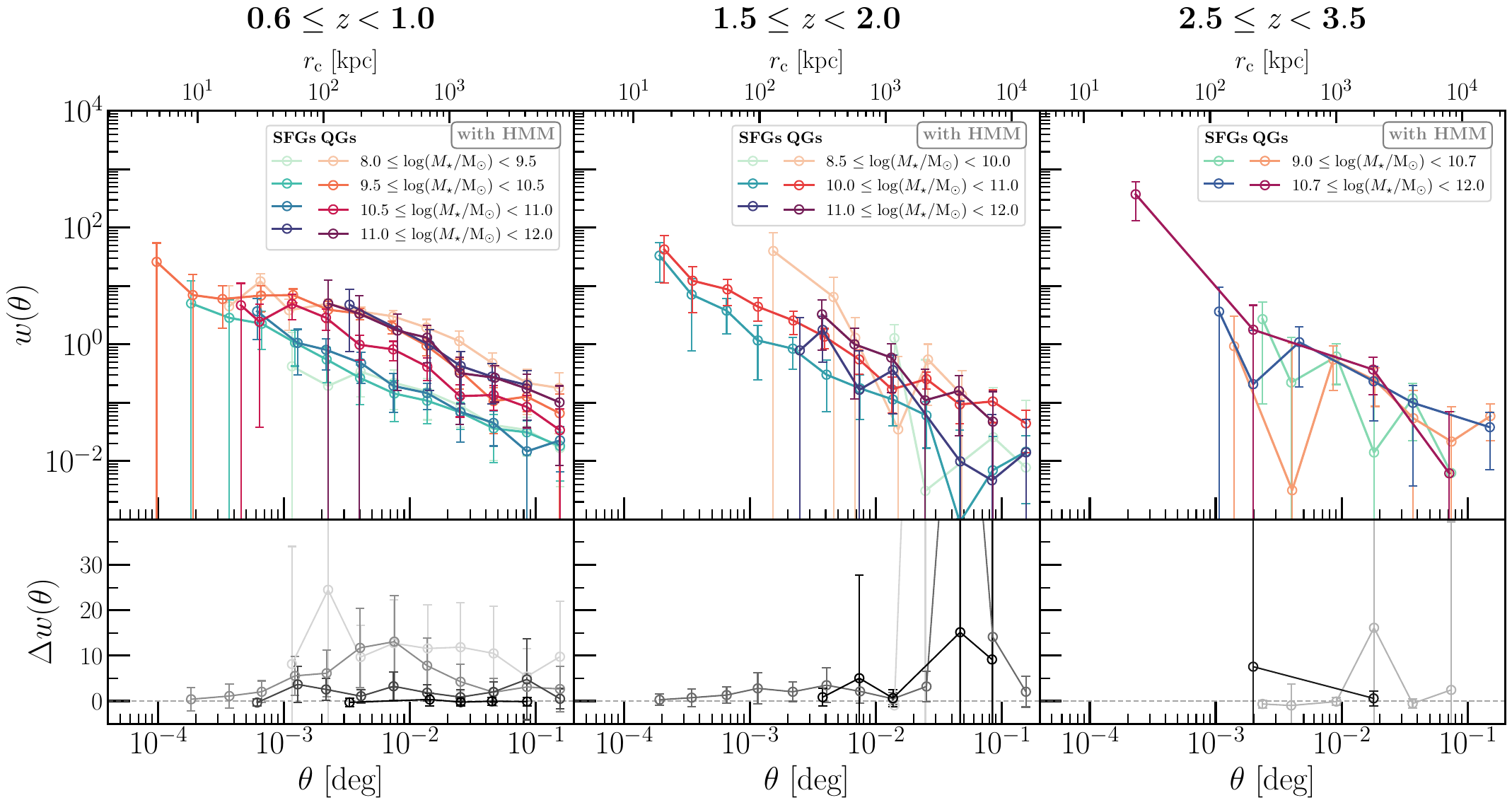}
    \caption{Similar to \reffig{fig:autocorr_ranges}, but after performing halo mass matching.}
    \label{app-fig:autocorr_ranges_HMM}
\end{figure}

\clearpage

\section{Cross-correlations of SFGs and QGs} \label{app:crosscorr_withoutHMM}

\begin{figure*}[h!]
    \centering
    \includegraphics[height=0.91\textheight]{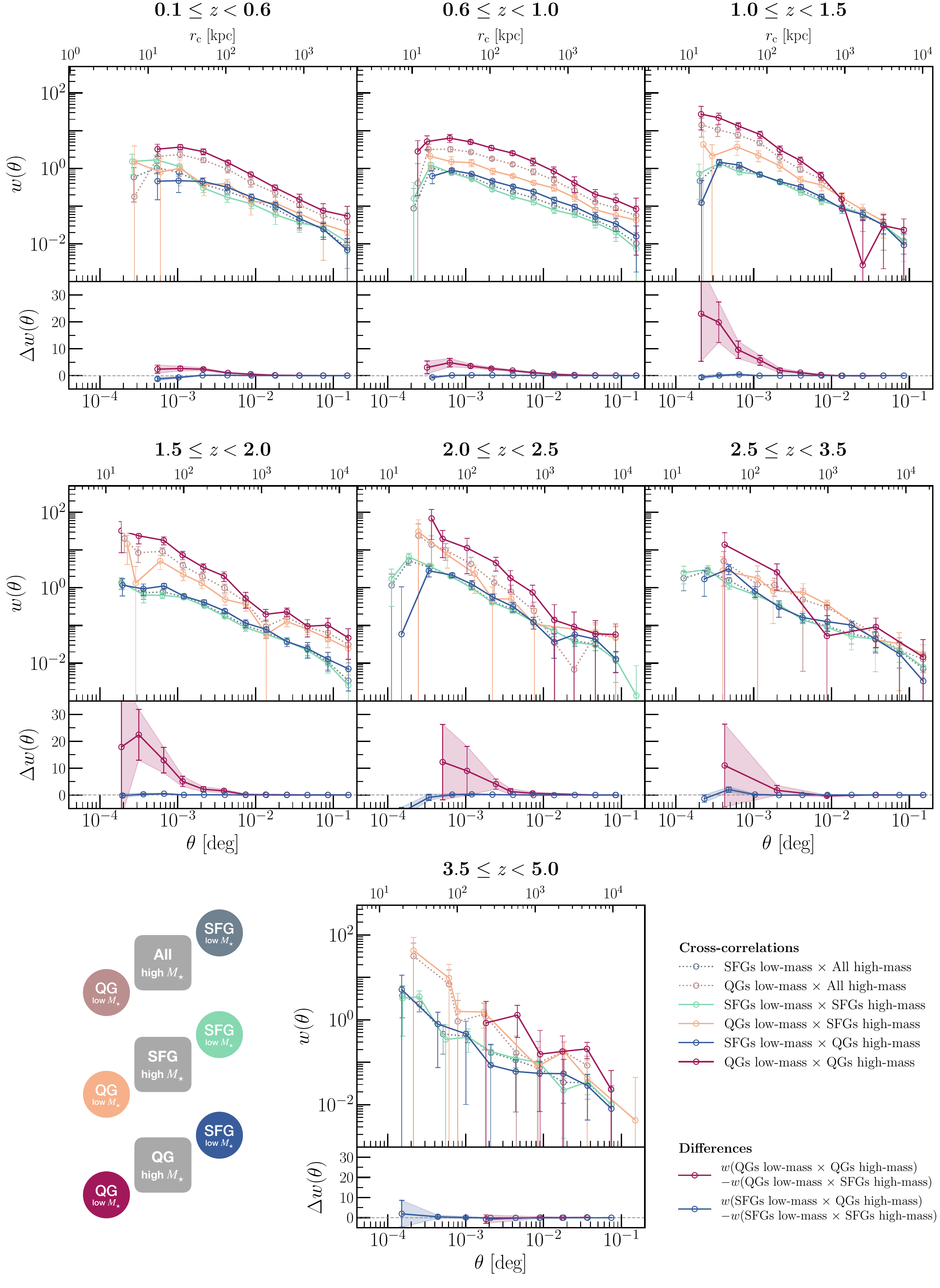}
    \caption{Similar to \reffig{fig:crosscorr_SM_HMM}, but without performing halo mass matching. We see a very significant conformity detection at the one-halo scales, up to $z = 2$ only.}
    \label{app-fig:crosscorr_SM}
\end{figure*}

\clearpage
\twocolumn

\section{Clustering of quiescent and star-forming galaxies in \textsc{UniverseMachine}} \label{app:autocorr_UM}

\begin{figure}[h!]
    \centering
    \includegraphics[width=0.68\columnwidth]{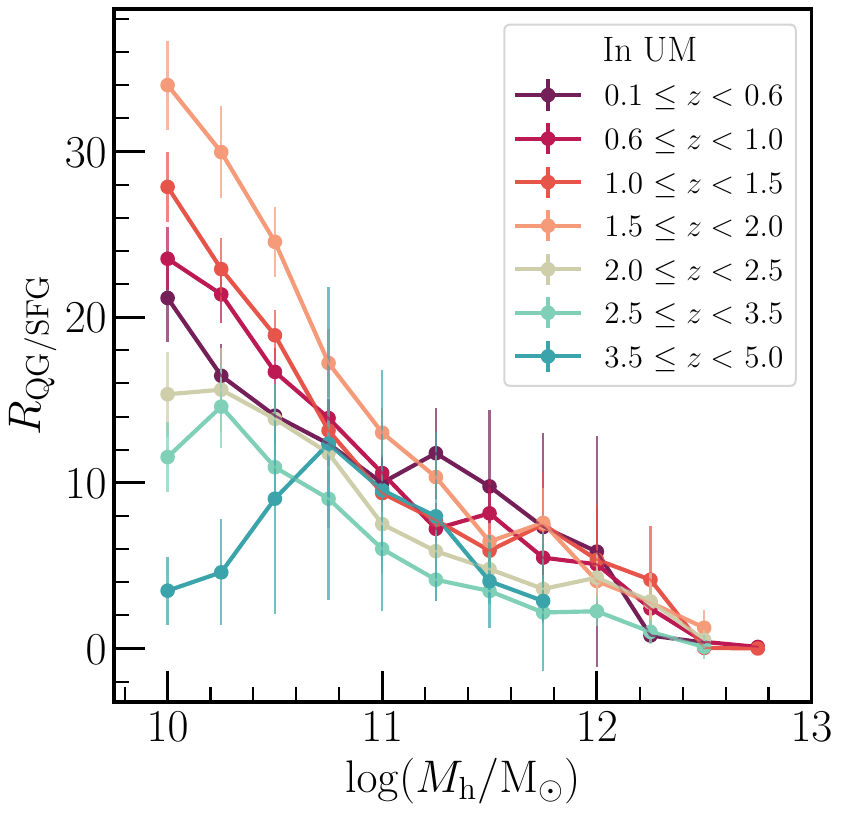}
    \caption{Ratio of clustering amplitudes $R_{\rm QG/SFG} = A_{\rm QG}/A_{\rm SFG}$ of \UM galaxies, as a function of halo mass and redshift, and for one-halo spatial scales.}
    \label{app-fig:summary_autocorr_SM}
\end{figure}

We did the same exercise as with COSMOS-Web galaxies presented in \refsec{sec:results}, but in \UM instead. In a similar manner to \reffig{fig:autocorr_sum}, we show the ratio between the clustering of QGs and that of SFGs, but within halo mass bins instead of stellar mass. This is presented for the one-halo regime in \reffig{app-fig:summary_autocorr_SM}. These results are consistent with our trends based on stellar mass: $R_{\rm QG/SFG}$ increases with decreasing halo mass, indicating that QGs in lower-mass halos become more and more clustered (most likely because they are satellites or reside in overdense regions). Beyond the critical halo mass $\log(M_{\rm h}/{\rm M}_\odot) \simeq 12$, SFGs and QGs appear clustered the same. Interestingly, the amplitude of $R_{\rm QG/SFG}$ is higher at cosmic noon than at lower redshift, an effect that could be attributed to the peak activity in clusters.

\section{Testing the limitations in detecting conformity} \label{app:limitations}

In this part, we test the two main limitations that could induce a wrong conformity detection: the control for halo mass between SFGs and QGs, and the central/satellite identification.

\begin{figure}[h!]
    \centering
    \includegraphics[width=0.95\columnwidth]{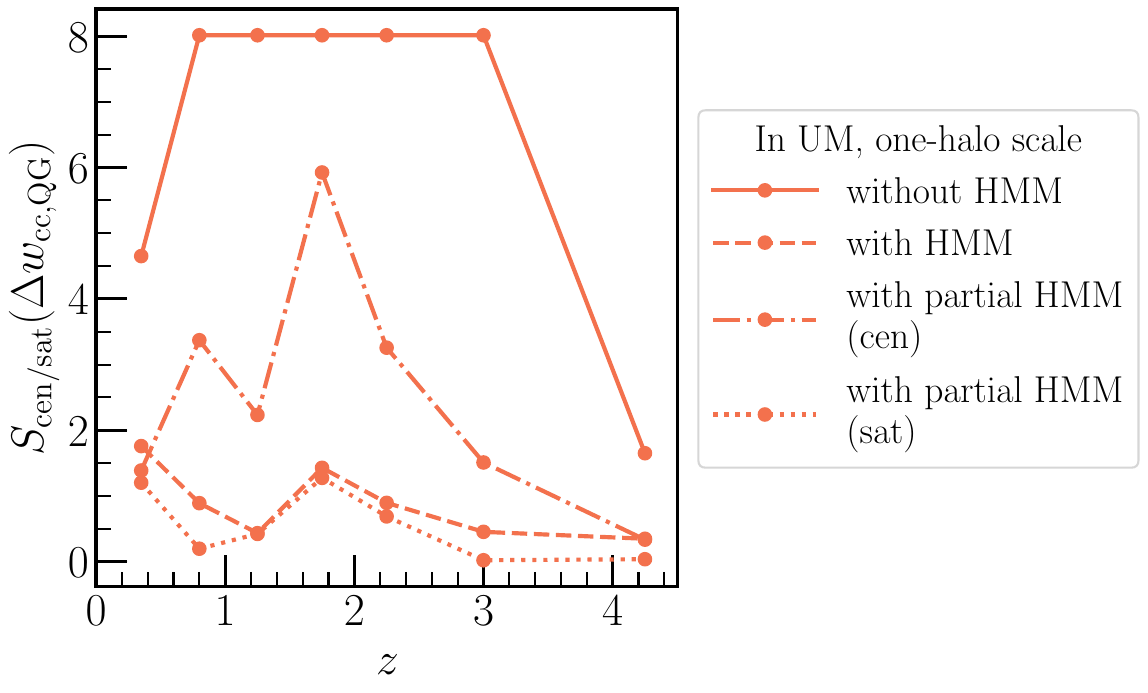}
    \caption{Detection significance of the one-halo conformity signal for quiescent satellites as a function of redshift, as defined in \refsec{sec:crosscorr} but here computed for \UM galaxies. This considers four cases that vary the halo mass-matching procedure. We note that $S$ is numerically capped at $8\sigma$ when the $p$-value equals 0.}
    \label{app-fig:limit_HMM}
\end{figure}

First, we investigate the impact of halo-mass matching (HMM) in \UM, where the true halo masses of galaxies are known. We measure the significance of one-halo conformity for quiescent satellites around quiescent centrals (compared to star-forming centrals), under four cases: (1) without matching the halo-mass distributions of SFGs and QGs; (2) with HMM; (3) with a partial HMM in which the 10\% of quiescent centrals residing in the most massive halos are omitted in the matching; and (4) with an analogous partial HMM but applied to quiescent satellites. As shown in \reffig{app-fig:limit_HMM}, applying HMM dramatically reduces the conformity signal, decreasing its significance from $>8\sigma$ to only $0.5\sigma$ at $z<3$. We note that some conformity is expected in \UM, as if conformity exists in the COSMOS field, the \UM model should have inherited it by design. But more importantly, we find that leaving just 10\% of QG centrals in the most massive halos unmatched increases the conformity significance by more than a factor of $2-4$, reaching $>2\sigma$ at $z<3$. This result demonstrates that the inferred amplitude of the conformity signal is highly sensitive to the accuracy of the halo-mass control, and that even small mismatches can produce a substantial artificial enhancement of the signal.

\begin{figure}[h!]
    \centering
    \includegraphics[width=0.95\columnwidth]{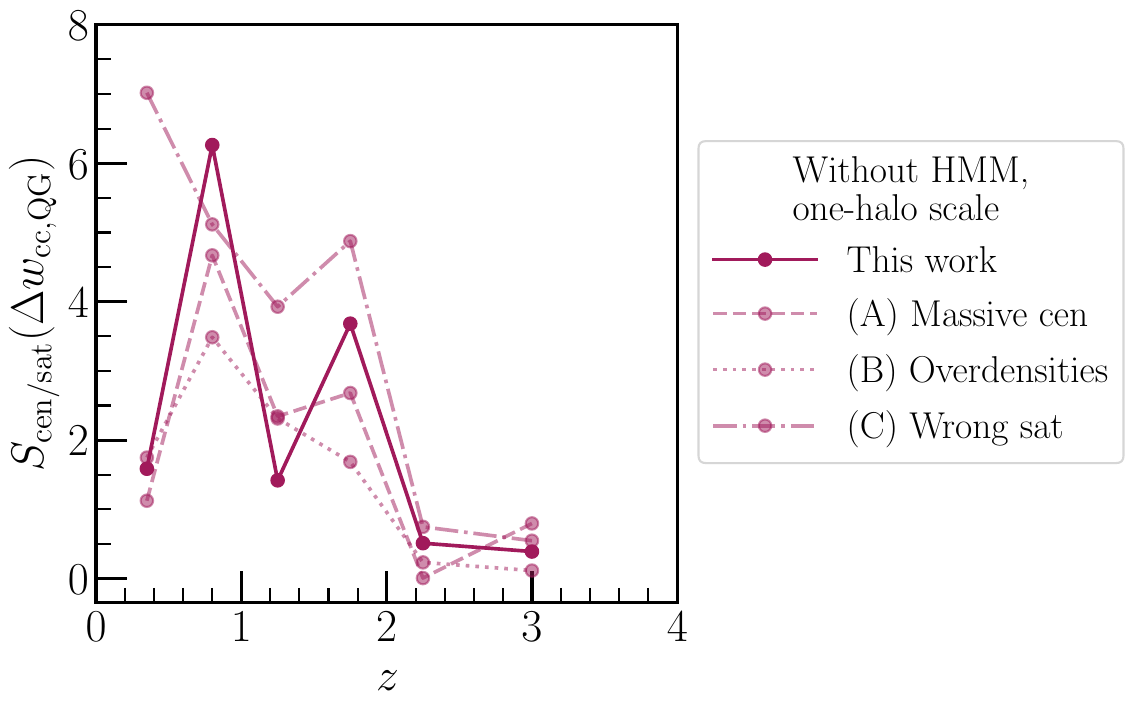}
    \caption{Detection significance of the one-halo conformity signal for quiescent satellites as a function of redshift, as defined in \refsec{sec:crosscorr}, for COSMOS-Web galaxies. This considers four cases that vary the central and satellite subsamples.}
    \label{app-fig:limit_censat}
\end{figure}

Second, we investigate the impact of central and satellite classification, since previous studies have suggested that conformity can arise solely from systematics in this process \citep[\eg][]{Calderon2018}, or be driven by a small number of low-mass centrals located near massive clusters \citep[\eg][]{Lacerna2018}. To assess this effect, we repeat the previous exercise in COSMOS-Web while modifying the central and satellite samples. \reffigure{app-fig:limit_censat} presents the conformity significance for four different selections: the fiducial sample; (A) a sample in which the 10\% most massive centrals are removed; (B) a sample in which the 20\% of centrals residing in the most overdense environments are removed; and (C) a sample in which the 10\% most massive satellites are reassigned as centrals. We find that the conformity signal is enhanced when massive satellites are misclassified as centrals, highlighting the measurement's sensitivity to central-satellite identification. Moreover, a substantial fraction of the signal appears to be driven by centrals located in the most overdense environments, since removing them reduces the significance by a factor of $\sim 2$ at $z<2$. This confirms that surveys covering small sky areas may be particularly susceptible to cosmic variance: fields containing rich clusters are more likely to detect an enhanced conformity signal than other regions.

\end{appendix}

\end{document}